%% file: mainvs.tex
\documentclass[conference]{IEEEtran}
\usepackage{url}
\usepackage{graphicx}
\usepackage{balance}
\usepackage{subfigure}
\usepackage{color}
\usepackage{times,amsmath,epsfig}
\usepackage{amssymb}
\usepackage{textcomp}
\usepackage[linesnumbered,algoruled,boxed,lined,noend]{algorithm2e}
\usepackage{cite}
\usepackage{todonotes} 

\newcommand{\yy}[1]{{\color{magenta} #1}\normalcolor}

\newcommand{\remove}[1]{}

\newcommand{\ignore}[1]{}

\usepackage{booktabs} 

\begin{document}
	
\title{MESSI: In-Memory Data Series Indexing}
%



\author{\IEEEauthorblockN{Botao Peng}
	\IEEEauthorblockA{LIPADE, Universit{\'e} de Paris\\
		botao.peng@parisdescartes.fr}
	\and
	\IEEEauthorblockN{Panagiota Fatourou}
	\IEEEauthorblockA{FORTH ICS \& Dept. of Comp. Science, Univ. of Crete\\
		faturu@csd.uoc.gr}
	\and
	\IEEEauthorblockN{Themis Palpanas}
	\IEEEauthorblockA{LIPADE, Universit{\'e} de Paris\\
		themis@mi.parisdescartes.fr}
}
\maketitle

\begin{abstract}
Data series similarity search is a core operation for several data series analysis applications across many different domains. 
However, the state-of-the-art techniques fail to deliver the time performance required for interactive exploration, 
or analysis of large data series collections.
In this work, we propose MESSI, the first data series index designed for in-memory operation on modern hardware.
Our index takes advantage of the modern hardware parallelization opportunities (i.e., SIMD instructions, multi-core and multi-socket architectures), in order to accelerate both index construction and similarity search processing times. 
Moreover, it benefits from a careful design in the setup and coordination of the parallel workers and data structures, so that it maximizes its performance for in-memory operations.
Our experiments with synthetic and real datasets demonstrate that overall MESSI is up to 4x faster at index construction, and up to 11x faster at query answering than the state-of-the-art parallel approach. 
MESSI is the first to answer exact similarity search queries on 100GB datasets in $\sim$50msec (30-75msec across diverse datasets), which enables real-time, interactive data exploration on very large data series collections.
\end{abstract}

%




\input{introduction.tex}

\input{preliminaries.tex}

\input{indexcreation.tex}
\input{queryanswering.tex}

\input{experiments.tex}

\input{relatedwork.tex}

\input{conclusions.tex}
\bibliographystyle{IEEEtran}
\bibliography{parisinmemory}

\end{document}

%% file: introduction.tex
\section{Introduction}

\noindent{\bf{[Motivation]}}
Several applications across many diverse domains, such as in finance, astrophysics, neuroscience, engineering, multimedia, 
and others~\cite{DBLP:journals/sigmod/Palpanas15, fulfillingtheneed,Palpanas2019}, continuously produce big collections of data series\footnote{A data series, or data sequence, 
	is an ordered sequence of data points. If the ordering dimension is time then we talk 
	about time series, though, series can be ordered over other measures. 
	(e.g., angle in astronomical radial profiles, frequency in infrared spectroscopy, mass in mass spectroscopy, position in genome sequences, etc.).}
which need to be processed and analyzed.
The most common type of query that different analysis applications need to answer on these collections of data series 
is similarity search~\cite{DBLP:journals/sigmod/Palpanas15, lernaeanhydra, lernaeanhydra2}.

The continued increase in the rate and volume 
of data series production
renders existing data series indexing technologies inadequate. 
For example, ADS+~\cite{zoumpatianos2016ads}, the state-of-the-art sequential (i.e., non-parallel) indexing technique, 
requires more than 2min to answer exactly a single 1-NN (Nearest Neighbor) query on a (moderately sized) 100GB sequence dataset.
For this reason, a disk-based data series parallel indexing scheme, called ParIS, 
was recently designed~\cite{peng2018paris} to take advantage 
of modern hardware parallelization. 
ParIS effectively exploits the parallelism capabilities provided by multi-core 
and multi-socket architectures, 
and the Single Instruction Multiple Data (SIMD) capabilities 
of modern CPUs.
In terms of query answering, experiments showed that ParIS is more than $1$ order of magnitude faster 
than ADS+, and more than $3$ orders of magnitude faster
than the optimized serial scan method. 

Still, ParIS is designed for disk-resident data and therefore
its performance is dominated by the I/O costs it encounters. 
For instance, ParIS answers a 1-NN (Nearest Neighbor) exact query on a 100GB dataset in 15sec, 
which is above the limit for keeping the user's attention (i.e., 10sec), 
let alone for supporting interactivity in the analysis process (i.e., 100msec)~\cite{Fekete:2016}.


\noindent{\bf{[Application Scenario]}}
In this work, we focus on designing an efficient parallel indexing and query answering scheme 
for \emph{in-memory} data series processing. Our work is motivated and inspired by the following
real scenario. Airbus\footnote{\scriptsize\url{http://www.airbus.com/}}, 
currently stores petabytes of data series, describing the behavior 
over time of various aircraft components (e.g., the vibrations of the bearings in the engines), 
as well as that of pilots (e.g., the way they maneuver the plane through the fly-by-wire system)~\cite{Airbus}. 
The experts need to access these data in order to run different analytics algorithms. 
However, these algorithms usually operate on a subset of the data (e.g., only the data relevant to landings 
from Air France pilots), which fit in memory. 
Therefore, in order to perform complex analytics operations (such as searching for similar patterns, or classification) fast, in-memory data series indices must be built for efficient data series query processing. 
Consequently, the time performance of both index creation and query answering become important factors in this process.

\noindent{\bf{[MESSI Approach]}}
We present MESSI, the {\em first} in-MEmory data SerieS Index, which
incorporates the state-of-the-art techniques in sequence indexing. 
MESSI effectively uses multi-core and multi-socket architectures 
in order to concurrently execute the computations needed for both index construction 
and query answering and it exploits SIMD. 
%
More importantly though, MESSI features redesigned algorithms that lead to a further $\sim$4x speedup in index construction time, in comparison to an in-memory version of ParIS.
Furthermore, MESSI answers exact 1-NN queries on 100GB datasets 6-11x faster than ParIS across the datasets we tested, achieving for the first time interactive exact query answering times, at $\sim$50msec.

When building ParIS, the design decisions were heavily influenced by the fact that 
the cost was mainly I/O bounded. 
Since MESSI copes with in-memory data series, no CPU cost can be hidden under I/O.
Therefore, MESSI required more careful design choices 
and coordination of the parallel workers when accessing
the required data structures, in order to improve its performance. 
This led to the development of a more subtle design for the construction of the index 
and on the development of new algorithms for answering similarity search queries on this index. 

For query answering in particular, we showed that adaptations of alternative solutions,
which have proven to perform the best in other settings (i.e., disk-resident data~\cite{peng2018paris}), 
are not optimal in our case, and we designed a novel solution that achieves a good balance 
between the amount of communication among the parallel worker threads, and the effectiveness of each individual worker.
For instance, the new scheme uses concurrent priority queues for storing the data series 
that cannot be pruned, and for processing these series in order, starting from those whose iSAX representations have the smallest distance
to the iSAX representation of the query data series. 
In this way, the parallel query answering threads achieve better pruning on the data series they process.
Moreover, the new scheme uses the index tree to decide which data series to insert 
into the priority queues for further processing. In this way, the number of distance calculations performed
between the iSAX summaries of the query and data series is significantly reduced (ParIS 
performs this calculation for all data series in the collection).
We also experimented with several designs
for reducing the synchronization cost among different workers that access 
the priority queues and for achieving load balancing.
We ended up with a scheme where 
workers use radomization to choose the priority queues they will work on.
Consequently, MESSI answers exact 1-NN queries on 100GB datasets within 30-70msec across diverse synthetic and real datasets.

The index construction phase of MESSI differentiates from ParIS in several ways.
For instance, ParIS was using a number of buffers to temporarily store pointers to the iSAX summaries
of the raw data series before constructing the tree index~\cite{peng2018paris}.
MESSI allocates smaller such buffers per thread and stores in them the iSAX summaries themselves.
In this way, it completely eliminates the synchronization cost in accessing the iSAX buffers.
To achieve load balancing, MESSI splits the array storing the raw data series into small blocks, and assigns blocks to threads
in a round-robin fashion. 
We applied the same technique when assigning to threads the buffers containing the iSAX summary of the data series.
Overall, the new design and algorithms of MESSI led to $\sim$4x improvement in index construction time when compared to ParIS.

\noindent{\bf [Contributions]} 
Our contributions are summarized as follows. 
\begin{itemize}
\item
We propose MESSI, the first in-memory data series index designed for modern hardware, 
which can answer similarity search queries in a highly efficient manner.
\item
We implement a novel, tree-based exact query answering algorithm, 
which minimizes the number of required distance calculations 
(both lower bound distance calculations for pruning true negatives, 
and real distance calculations for pruning false positives).
\item
We also design an index construction algorithm that effectively balances 
the workload among the index creation workers by using 
a parallel-friendly index framework with low synchronization cost.
\item
We conduct an experimental evaluation with several synthetic and real datasets, 
which demonstrates the efficiency of the proposed solution. 
The results show that MESSI is up to 4.2x faster at index construction and up to 11.2x faster at query answering than the state-of-the-art parallel index-based competitor, up to 109x faster at query answering than the state-of-the-art parallel serial scan algorithm, and thus can significantly reduce the execution time of complex analytics algorithms (e.g., \emph{k-NN} classification). 
\end{itemize}

%% file: preliminaries.tex
\section{Preliminaries}
\label{sec:prelim}

We now provide some necessary definitions, and introduce the related work on state-of-the-art data series indexing.

\subsection{Data Series and Similarity Search}

\noindent{\bf [Data Series]}
A data series, $S=\{p_1, ..., p_n\}$, is defined as a sequence of points, 
where each point $p_i=(v_i,t_i)$, $1 \le i \le n$\yy{,} 
is associated to a real value $v_i$ and a position $t_i$.
The position corresponds to the order of this value in the sequence. 
We call $n$ the \emph{size}, or \emph{length} of the data series.
We note that all the discussions in this paper are applicable to high-dimensional vectors, in general.

\noindent{\bf [Similarity Search]}
Analysts perform a wide range of data mining tasks on data series 
including clustering~\cite{rakthanmanon2011},
classification and deviation detection~\cite{Shieh2009,Shandola2009}, 
and frequent pattern mining~\cite{DBLP:journals/datamine/MueenKZCWS11}.
Existing algorithms for executing these tasks rely on performing fast similarity search 
across the different series.
Thus, efficiently processing nearest neighbor (NN) queries is crucial 
for speeding up the above tasks.
NN queries are formally defined as follows: given a query series $S_q$ of length $n$,  
and a data series collection $\mathcal{S}$ of sequences of the same length, $n$, 
we want to identify the series $S_c \in \mathcal{S}$ 
that has the smallest distance to $S_q$ among all the series in the collection $\mathcal{S}$.
(In the case of streaming series, we first create subsequences of length $n$ using a sliding window, and then index those.)

Common distance measures for comparing data series are Euclidean Distance (ED)~\cite{Agrawal1993} 
and dynamic time warping (DTW)~\cite{rakthanmanon2012searching}.
While DTW is better for most data mining tasks, 
the error rate using ED converges to that of DTW 
as the dataset size grows~\cite{shieh2008sax}.
Therefore, data series indexes for massive datasets use ED 
as a distance metric~\cite{shieh2008sax,rakthanmanon2012searching,wang2013data,isax2plus,zoumpatianos2016ads}, 
though simple modifications can be applied to make them compatible with DTW~\cite{shieh2008sax}.
Euclidean distance is computed as the sum of distances between the pairs of corresponding points in the two sequences.
Note that minimizing ED on z-normalized data (i.e., a series whose values have mean 0 and standard deviation 1) is equivalent to maximizing their Pearson's correlation coefficient~\cite{MueenNL10}.

\noindent{\bf [Distance calculation in SIMD]}
Single-Instruction Multiple-Data (SIMD) refers to a parallel architecture that allows the execution of the same operation on multiple data simultaneously~\cite{lomont2011introduction}. 
Using SIMD, we can reduce the latency of an operation, because the corresponding instructions are fetched once, and then applied in parallel to multiple data.
All 
modern CPUs support 256-bit wide SIMD vectors, 
which means that certain floating point (or other 32-bit data) computations can be up to 8 times faster when executed using SIMD.
%

In the data series context, SIMD has been employed for the computation of the Euclidean distance functions~\cite{tang2016exploit}, as well as in the ParIS index, for the conditional branch calculations during the computation of the lower bound distances~\cite{peng2018paris}.


\begin{figure}[tb]

\begin{minipage}[b]{0.4\columnwidth}
\subfigure[raw data series\label{fig:saxa}] {
	\hspace{-1em}
	\includegraphics[page=1,width=0.95\columnwidth]{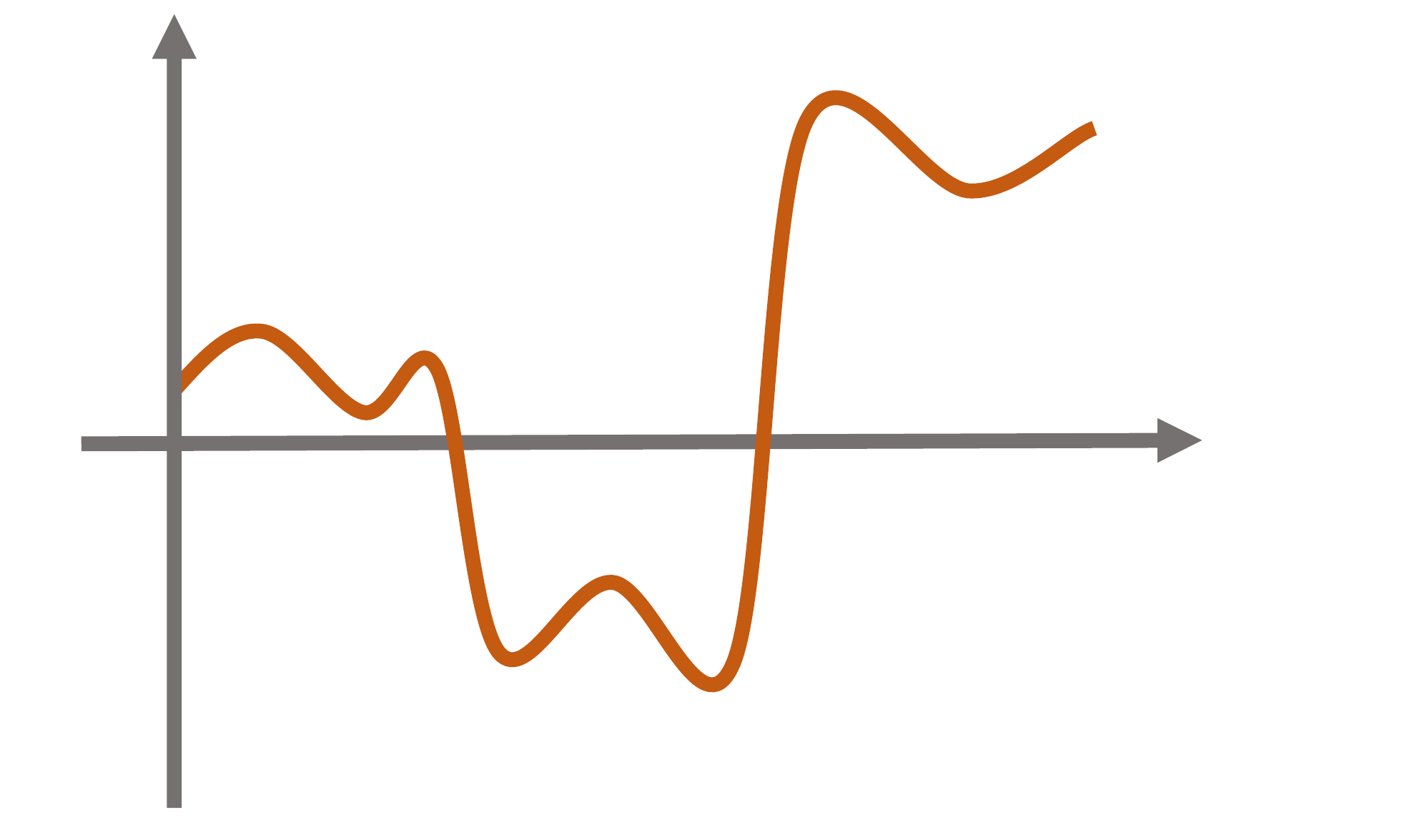}
}

\subfigure[PAA representation\label{fig:saxb}] {
	\hspace{-1em}
	\includegraphics[page=2,width=0.95\columnwidth]{picture/sax2}
}
\subfigure[iSAX representation\label{fig:saxc}]{
	\hspace{-1em}
	\includegraphics[page=3,width=0.95\columnwidth]{picture/sax2}
}
\end{minipage}
\hspace*{-0.2cm}
\subfigure[ParIS index\label{fig:ads}] {
	\hspace{-1em}
	\includegraphics[width=0.62\columnwidth]{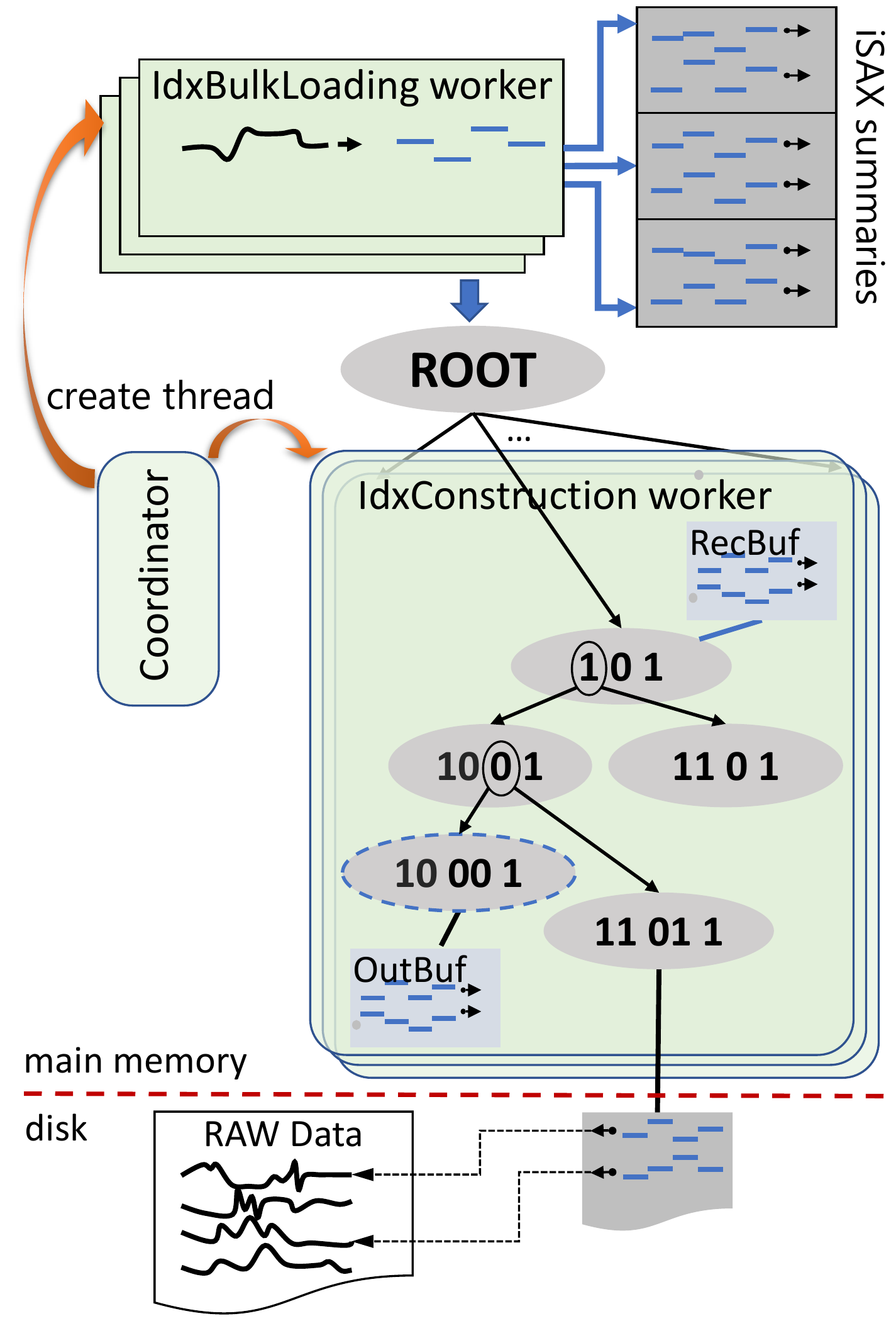}
}
\caption{The iSAX representation, and the ParIS index}
\end{figure}

\subsection{iSAX Representation and the ParIS Index}

\noindent{\bf [iSAX Representation]}
The iSAX representation (or summary) is based on the 
Piecewise Aggregate Approximation (PAA) representation~\cite{keogh2001dimensionality}, which divides the data series in segments of equal length, and uses the mean value of the points in each segment in order to summarize a data series. 
Figure~\ref{fig:saxb} depicts an example of PAA representation with three segments (depicted with the black horizontal lines), for the data series depicted in Figure~\ref{fig:saxa}. 
Based on PAA, the indexable Symbolic Aggregate approXimation (iSAX) representation was proposed~\cite{shieh2008sax} (and later used in several different data series indices~\cite{Shieh2009, zoumpatianos2016ads, DBLP:journals/pvldb/KondylakisDZP18, peng2018paris, ulissevldb}).
This method first divides the (y-axis) space in different regions, and assigns a bit-wise symbol to each region.
In practice, the number of symbols is small: 
iSAX achieves very good approximations with as few as 256 symbols, the maximum alphabet cardinality, $|alphabet|$, which can be represented by eight bits~\cite{isax2plus}.
It then represents each segment $w$ of the series
with the symbol of the region the PAA falls into, forming the word $10_2 00_2 11_2$ shown in Figure~\ref{fig:saxc} (subscripts denote the number of bits used to represent the symbol of each segment). 



\noindent{\bf [ParIS Index]}
Based on the iSAX representation, the state-of-the-art ParIS index was developed~\cite{peng2018paris}, which proposed techniques and algorithms specifically designed for modern hardware and disk-based data. 
ParIS makes use of variable cardinalities for the iSAX summaries (i.e., variable degrees of precision for the symbol of each segment) 
in order to build a hierarchical tree index (see Figure~\ref{fig:ads}), consisting of three types of nodes:
(i) the root node points to several children nodes, $2^w$ in the worst case (when the series in the collection 
cover all possible iSAX summaries); (ii) each inner node contains the iSAX summary of all the series
below it, and has two children; and (iii) each leaf node contains the iSAX summaries of all the series inside it, and pointers to the raw data (in order to be able to prune false positives and produce exact, correct answers), which reside on disk. 
When the number of series in a leaf node becomes greater than the maximum leaf capacity, the leaf splits: 
it becomes an inner node and creates two new leaves, by increasing the cardinality of the iSAX summary 
of one of the segments (the one that will result in the most balanced split of the contents of the node 
to its two new children~\cite{isax2plus,zoumpatianos2016ads}). 
The two refined iSAX summaries (new bit set to \textit{0} and \textit{1}) are assigned to the two new leaves. 
In our example, the series of Figure~\ref{fig:saxc} will be placed in the outlined node of the index (Figure~\ref{fig:ads}).
Note that we define the distance of a query series to a node as the distance between the query (raw values, or iSAX summary) and the iSAX summary of the node.

In the index construction phase (see Figure~\ref{fig:ads}), ParIS uses a coordinator worker that reads raw data series from disk and transfers them into 
a raw data buffer in memory. 
A number of index bulk loading workers compute the iSAX summaries of these series, 
and insert $<$iSAX summary, file position$>$ pairs in an array. 
They also insert a pointer to the appropriate element of this array 
in the receiving buffer of the corresponding subtree of the index root.
When main memory is exhausted, the coordinator worker creates a number of index construction worker threads, each one assigned to one subtree of the root and responsible for further building that subtree (by processing the iSAX summaries stored in the coresponding receiving buffer). 
This process results in each iSAX summary being moved to the output buffer of the leaf it belongs to. 
When all iSAX summaries in the receiving buffer of an index construction worker have been processed, the output buffers of all leaves in that subtree are flushed to disk. 

For query answering, ParIS offers a parallel implementation of the SIMS exact search  algorithm~\cite{zoumpatianos2016ads}. 
It first computes an approximate answer by calculating the real distance between the query and the best candidate series, 
which is in the leaf with the smallest lower bound distance to the query. 
ParIS uses the index tree only for computing this approximate answer.
Then, a number of lower bound calculation workers compute the lower bound distances 
between the query and the iSAX summary of each data series in the dataset, which are stored in the \emph{SAX array},
and prune the series whose lower bound distance is larger than the approximate real distance computed earlier.
The data series that are not pruned, are stored in a candidate list for further processing.
Subsequently, a number of real distance calculation workers operate on different parts of this array
to compute the real distances between the query and the series stored in it
(for which the raw values need to be read from disk).
For details see~\cite{peng2018paris}.

In the in-memory version of ParIS, 
the raw data series are stored in an in-memory array.
Thus, there is no need for a coordinator worker. 
The bulk loading workers now operate directly on this array (split to as many chunks as the workers). 
In the rest of the paper, we use ParIS to refer to this in-memory version of the algorithm.

%% file: indexcreation.tex
\section{The MESSI Solution}
\label{sec:parmis}
\begin{figure}[tb]
	\centering
	\hspace*{-0.3cm}
	\includegraphics[page=1,width=1.05\columnwidth]{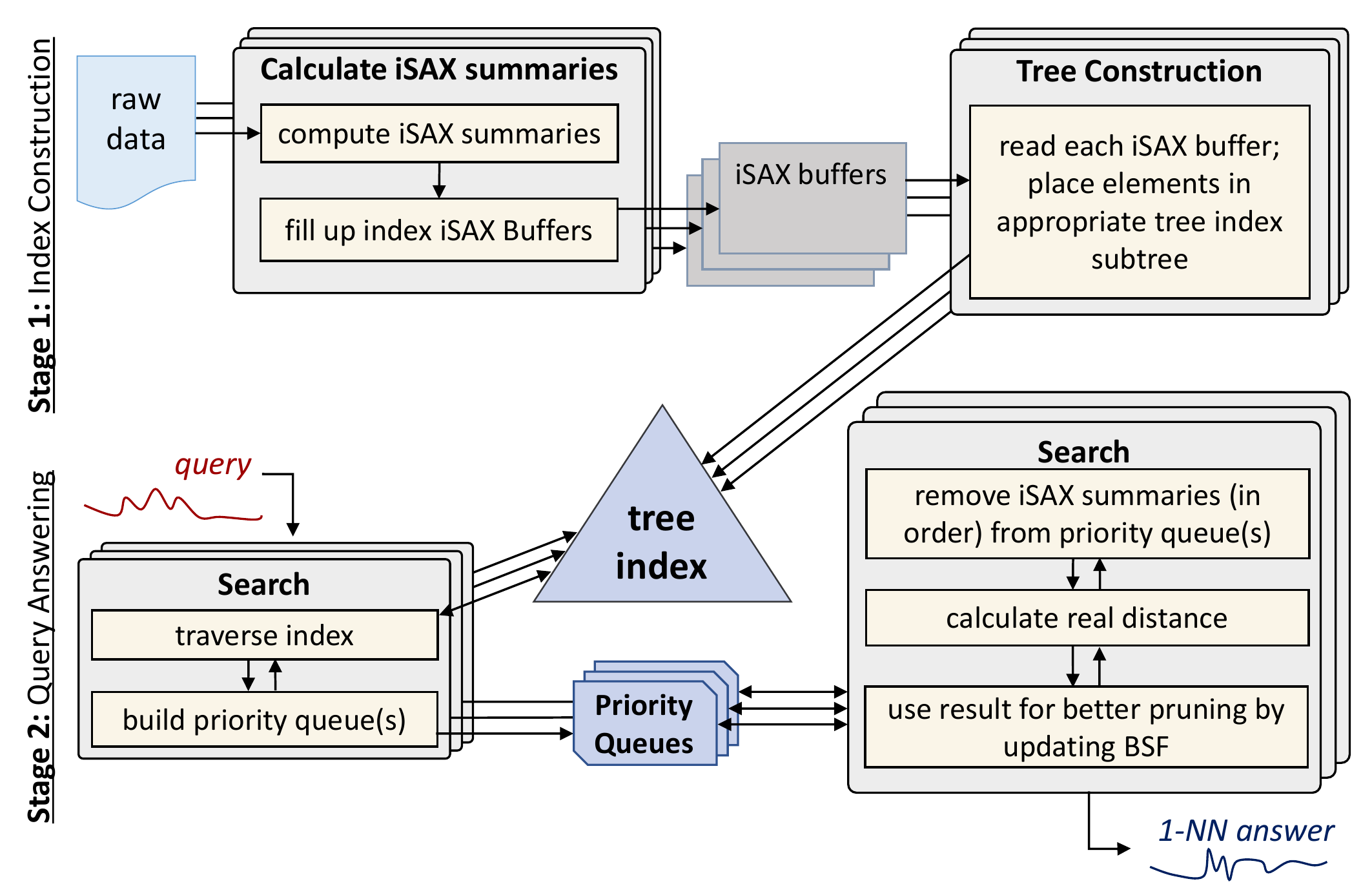}
	\caption{MESSI index construction and query answering}
	\label{fig:workflow}
\end{figure}

Figure~\ref{fig:workflow} depicts the MESSI index construction and query answering pipeline.
The raw data are stored in memory into an array, called $RawData$. 
This array is split into a predetermined number of chunks. 
A number, $N_w$, of {\em index worker} threads process the chunks
to calculate the iSAX summaries of the raw data series they store.
The number of chunks is not necessarily the same as $N_w$.
Chunks are assigned to index workers the one after the other (using Fetch\&Inc).
Based on the iSAX representation, we can figure out 
in which subtree of the index tree an iSAX summary will be stored.
A number of {\em iSAX buffers}, one for each root subtree of the index tree, 
contain the iSAX summaries to be stored in that subtree. 

Each index worker stores the iSAX summaries it computes in the appropriate iSAX buffers. 
To reduce synchronization cost, 
each iSAX buffer is split into parts and each worker works on its own part\footnote{
	We have also tried an alternative technique where each buffer 
	was protected by a lock and many threads were accessing each buffer.
        However, this resulted in worse performance due to 
	the encountered contention in accessing the iSAX buffers. 
	}.
The number of iSAX buffers is usually a few tens of thousands and at most $2^w$, 
where $w$ is the number of segments in the iSAX summaries 
of each data series ($w$ is fixed to $16$ in this paper, 
as in previous studies~\cite{zoumpatianos2016ads,peng2018paris}).

When the iSAX summaries for all raw data series have been computed, 
the index workers proceed in the constuction of the tree index. 
Each worker is assigned an iSAX buffer to work on (this is done
again using Fetch\&Inc). 
Each worker reads the data stored in (all parts of) 
its assigned buffer and builds the corresponding index subtree. 
Therefore, all index workers process distinct subtrees of the index, 
and can work in parallel and independently from one another, with no need for synchronization\footnote{
Parallelizing the processing inside each one of the index root subtrees would require 
a lot of synchronization due to node splitting. 
	}.
When an index worker finishes with the current iSAX buffer it works on, 
it continues with the next iSAX buffer that has not yet been processed.

When the series in all iSAX buffers have been processed, 
the tree index has been built and can be used to answer similarity search queries, as depicted in 
the query answering phase of Fig.~\ref{fig:workflow}. 
To answer a query, we first perform a search for the query iSAX summary in the tree index.
This returns a leaf whose iSAX summary has the closest distance
to the iSAX summary of the query.
We calculate the real distance of the (raw) data series pointed to by the elements
of this leaf to the query series, and store the minimum of these distances 
into a shared variable, called BSF (Best-So-Far). 
Then, the index workers start traversing the index subtrees
(the one after the other) using BSF to decide 
which subtrees will be pruned. 
The leaves of the subtrees that cannot be pruned are placed 
into (a fixed number of) minimum priority queues, using the lower bound distance between the raw values of the query series and the iSAX summary of the leaf node, in order to be further examined. 
Each thread inserts elements in the priority queues in a round-robin fashion
so that load balancing is achieved (i.e., all queues contain about 
the same number of elements). 

As soon as the necessary elements have been placed in the priority queues,
each index worker chooses a priority queue to work on, and repeatedly calls DeleteMin() on it to get a leaf node, on which it performs the following operations. 
It first checks whether the lower bound distance stored in the priority queue is larger than the current BSF: if it is then we are certain that the leaf node does not contain any series that can be part of the answer, and we can prune it; otherwise, the worker needs to examine the series contained in the leaf node, by first computing lower bound distances using the iSAX summaries, and if necessary also the real distances using the raw values. 
During this process, we may discover a series with a smaller distance to the query, in which case we also update the BSF.
When a worker reaches a node whose distance
is bigger than the BSF, it gives up this priority queue
and starts working on another, because it is certain
that all the other elements in the abandoned queue have an even higher distance to the query series. 
This process is repeated until
all priority queues have been processed.
During this process, the value of BSF is updated to always 
reflect the minimum distance seen so far. 
At the end of the calculation,
the value of BSF is returned as the query answer. 

Note that, similarly to ParIS, MESSI uses SIMD (Single-Instruction Multiple-Data) 
for calculating the distances of both, the index iSAX summaries 
from the query iSAX summary ({\em lower bound distance calculations}), 
and the raw data series from the query data series ({\em real distance calculations})~\cite{peng2018paris}.

\subsection{Index Construction}
Algorithm~\ref{creatindex-bp} presents the pseudocode for the {\em initiator} thread.
The initiator creates $N_w$ index worker threads 
to execute the index construction phase (line~\ref{crb:worker}). 
As soon as these workers finish their execution, 
the initiator returns (line~\ref{crb:return}). 
We fix $N_w$ to be $24$ threads (Figure~\ref{fig:pRecBuf} in Section~\ref{sec:experiments} justifies this choice).
We assume that the $index$ variable is a structure (struct) containing the $RawData$ array,
all iSAX buffers, and a pointer to the root of the tree index.
Recall that MESSI splits $RawData$ into chunks of size $chunk\_size$. 
We assume that the size of $RawData$ is a multiple of
$chunk\_size$ (if not, standard padding techniques can be applied).

\begin{figure*}[tb]
	\centering
	\subfigure[CalculateiSAXSummaries\label{fig:inc2a}]{
		\includegraphics[page=1,width=0.71\columnwidth]{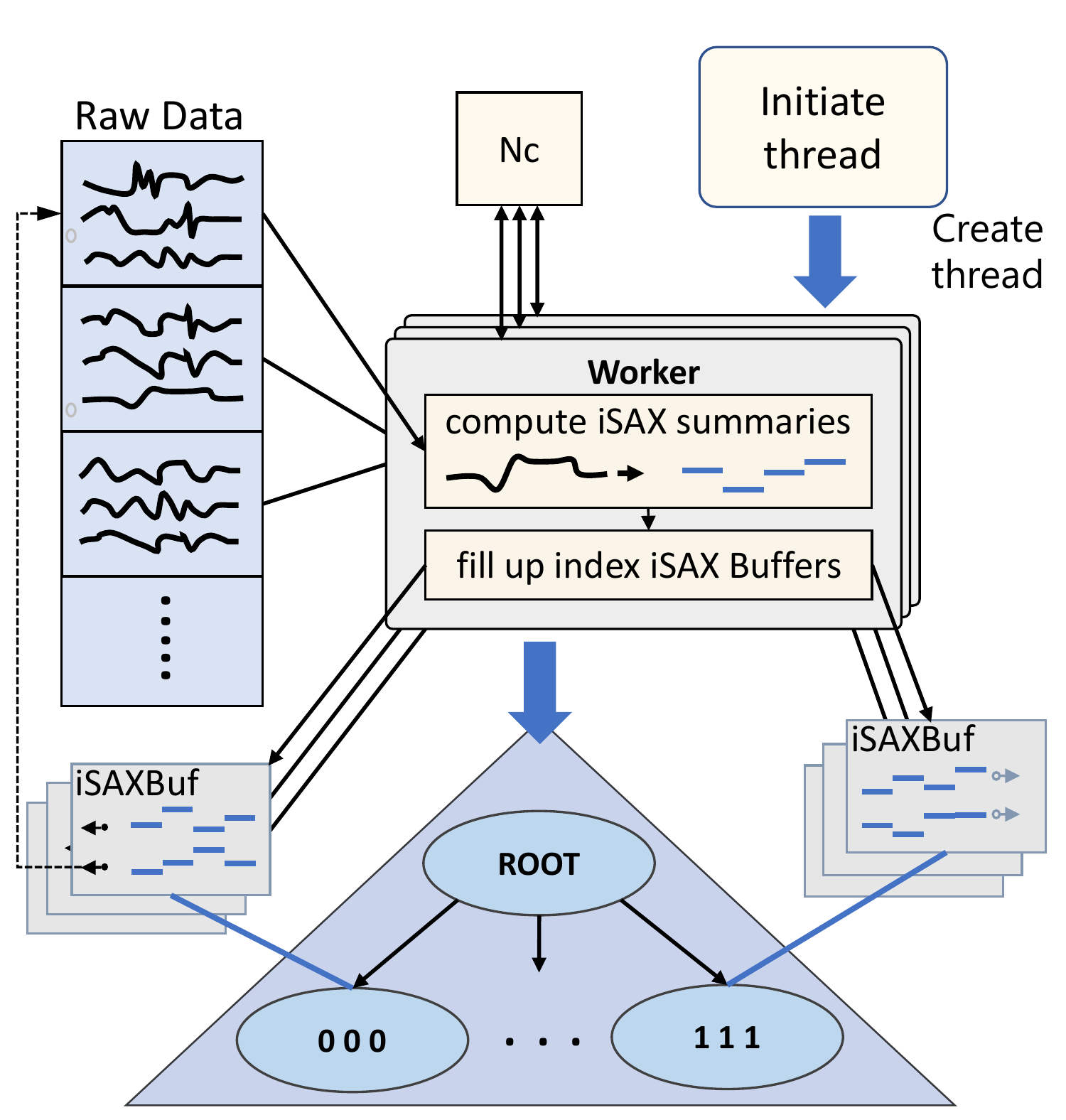}
	}
\hspace*{2cm}
	\subfigure[TreeConstruction\label{fig:inc2b}]{
		\includegraphics[page=1,width=0.73\columnwidth]{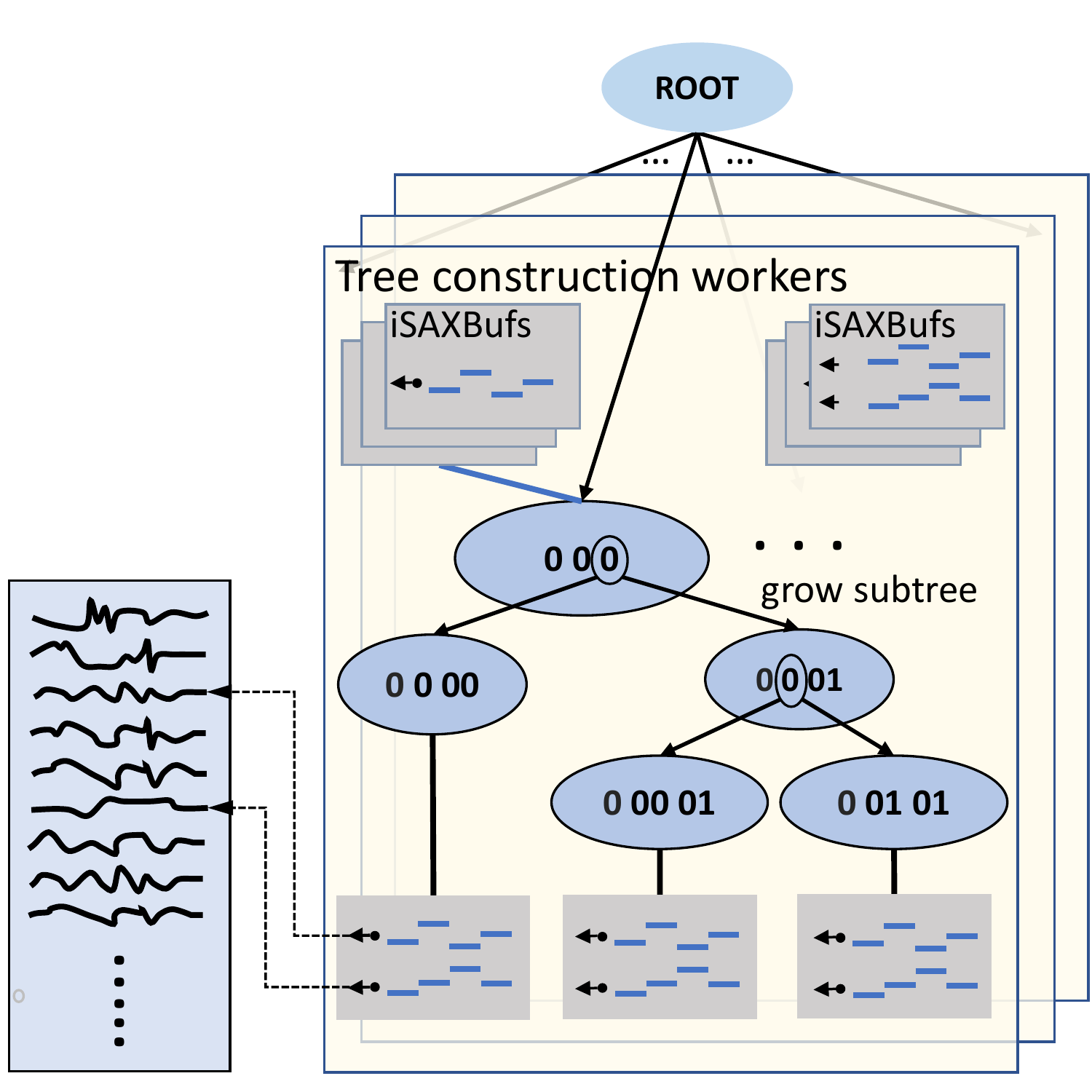}
	}
	\caption{Workflow and algorithms for MESSI index creation}
	\label{fig:inc2}
\end{figure*}

\begin{algorithm}[tb]
	{\footnotesize
		\SetAlgoLined
		\KwIn{\textbf{Index} $index$, \textbf{Integer} $N_w$, \textbf{Integer} $chunk\_size$	}	\vspace*{.1cm}
		\vspace*{.1cm}
		
		\For{$i$ $\leftarrow$ $0$ \emph{\KwTo} $N_w-1$} {
			create a thread to execute an instance of IndexWorker($index$, $chunk\_size$,$i$, $N_w$)\label{crb:worker}\; 		
		}
               wait for all these threads to finish their execution\;\label{crb:return}

	}
	\caption{$CreateIndex$}
	\label{creatindex-bp}
\end{algorithm}

\begin{algorithm}[tb]
	{\footnotesize
		\SetAlgoLined
		\KwIn{\textbf{Index} $index$, \textbf{Integer} $chunk\_size$, \textbf{Integer} $pid$, \textbf{Integer} $N_w$}
		\vspace*{.1cm}
		CalculateiSAXSummaries($index$, $chunk\_size$,$pid$)\;\label{iw:callbl}
		barrier to synchronize the IndexWorkers with one another\;\label{iw:barrier}
		TreeConstruction($index$, $N_w$)\;\label{iw:calltc}
		exit()\;
	}
\caption{$IndexWorker$}
\label{indexworker}
\end{algorithm}

The pseudocode for the index workers is in Algorithm~\ref{indexworker}. 
The workers first call the $CalculateiSAXSummaries$ function (line~\ref{iw:callbl})
to calculate the iSAX summaries of the raw data series and store
them in the appropriate iSAX buffers. 
As soon as the iSAX summaries of all the raw data series have been computed
(line~\ref{iw:barrier}), 
the workers call $TreeConstruction$  
to construct the index tree. 

The pseudocode of $CalculateiSAXSummaries$ is shown in Algorithm~\ref{bulkloading-bp} 
and is schematically illustrated in Figure~\ref{fig:inc2a}. 
Each index worker repeatedly does the following. 
It first performs a Fetch\&Inc to get assigned a chunk of raw data series to work on
(line~\ref{al2b:part}). Then, it calculates the offset in the $RawData$ array that this
chunk resides (line~\ref{al2b:initb}) and starts processing the relevant data series (line~\ref{al2b:for}). 
For each of them, it computes its iSAX summary by calling the ConvertToiSAX function (line~\ref{al2b:cover}),
and stores the result in the appropriate iSAX buffer of $index$ (lines~\ref{al2b:mask}-\ref{al2b:insert}). 
Recall that each iSAX buffer is split into $N_w$ parts, one for each thread;
thus, $index.iSAXbuffer$ is a two dimensional array. 

Each part of an iSAX buffer
is allocated dynamically when the first element to be stored in it 
is produced.  
The size of each part has an initial small value (5 series in this work, as we discuss in the experimental evaluation) 
and it is adjusted dynamically based on how many elements are inserted in it 
(by doubling its size each time). 

We note that we also tried a design of MESSI with no iSAX buffers, but this led to slower performance (due to the worse cache locality). 
Thus, we do not discuss this alternative further.

\begin{algorithm}[tb]
	{\footnotesize
		\SetAlgoLined
		\KwIn{\textbf{Index} $index$, \textbf{Integer} $chunk\_size$, \textbf{Integer} $pid$}
		\textbf{Shared integer} $F_c=0$\;  
		\vspace*{.1cm}
		\While{(TRUE)}
		{
			$b\leftarrow${\em Atomically} fetch and increment $F_c$\;\label{al2b:part}
			$b$ = $b*chunk\_size$\;\label{al2b:initb}
			\textbf{if} ($b \geq$ size of the $index.RawData$ array) \textbf{then} break \;
			\For{$j$ $\leftarrow$ $b$ \emph{\KwTo} $b + chunk\_size$\label{al2b:for}}
			{
				$isax$ = $ConvertToiSAX$($index.RawData[j]$)\;\label{al2b:cover}
    			        $\ell$ = find appropriate root subtree where $isax$ must be stored\;\label{al2b:mask}
				$index.iSAXbuf[\ell][pid] = \langle isax, j \rangle$\;\label{al2b:insert}
			}
		}
	}
\caption{$CalculateiSAXSummaries$}
\label{bulkloading-bp}
\end{algorithm}

\begin{algorithm}[tb]
	{\footnotesize
		\SetAlgoLined
		\KwIn{\textbf{Index} $index$, \textbf{Integer} $N_w$} 
		\vspace*{.1cm}
		\textbf{Shared integer} $F_{b}=0$\;  
		\vspace*{.1cm}
		\While{(TRUE) }
		{
			$b\leftarrow${\em Atomically} fetch and increment $F_{b}$\;\label{al3:gotnode}
			\textbf{if} ($b \geq 2^w$) \textbf{then} break \tcp*{\footnotesize{the root has at most $2^w$ children}} 
			\For{$j$ $\leftarrow$  $0$ \emph{\KwTo} $N_w$\label{conp:lay}}
			{
				\For{\textbf{{\em every}} $\langle isax, pos \rangle$ {\em pair} $\in index.iSAXbuf[b][j]$\label{conp:passts}}
				{
					$targetLeaf \leftarrow$ Leaf of $index$ tree to insert $\langle isax, pos \rangle$\;\label{conp:insertinleaf}
					\While{$targetLeaf$ {\em is full}}
					{
						SplitNode($targetLeaf$)\;\label{al3:split}
						$targetLeaf \leftarrow$ New leaf to insert $\langle isax, pos \rangle$\;
					}
					Insert $\langle isax, pos \rangle$ in $targetLeaf$\;\label{al3:output}
				}
			}
		}
	}
\caption{$TreeConstruction$}
\label{construction-p}
\end{algorithm}

As soon as the computation of the iSAX summaries is over, 
each index worker starts executing the $TreeConstruction$ function.
Algorithm~\ref{construction-p} shows the pseudocode for this function and 
Figure~\ref{fig:inc2b} schematically describes how it works.
In $TreeConstruction$, a worker repeatedly executes the following actions.
It accesses $F_b$ (using Fetch\&Inc) to get assigned an iSAX buffer to work on (line~\ref{al3:gotnode}). 
Then, it traverses all parts of the assigned buffer (lines~\ref{conp:lay}-\ref{conp:passts}) 
and inserts every pair $\langle \mbox{iSAX summary}, \mbox{pointer to relevant data series} \rangle$ 
stored there in the index tree (line~\ref{conp:insertinleaf}-\ref{al3:output}). 
Recall that the iSAX summaries contained
in the same iSAX buffer will be stored in the same subtree of the index tree.
So, no synchronization is needed among the index workers during this process. 
If a tree worker finishes its work on a subtree, 
a new iSAX buffer is (repeatedly) assigned to it, until all iSAX buffers have been processed. 

%% file: queryanswering.tex
\subsection{Query Answering}

\begin{algorithm}[tb]
	{\footnotesize	
		\SetAlgoLined
		\textbf{Shared float} $BSF$\;
		\KwIn{\textbf{QuerySeries} $QDS$, \textbf{Index} $index$, \textbf{Integer} $N_q$ }
		QDS\_iSAX = calculate iSAX summary for QDS\;\label{eshq:calcuqisax}
		BSF = approxSearch($QDS\_iSAX$, $index$)\;\label{eshq:appro}
		\For{$i$ $\leftarrow$ $0$ \emph{\KwTo} $N_q-1$\label{eshq:scq}} {
			$queue[i]$ = Initialize the $i$th priority queue\;
		}\label{eshq:ecq}

		\For{$i$ $\leftarrow$ $0$ \emph{\KwTo} $N_s-1$\label{eshq:scw}} {
			create a thread to execute an instance of  		
			SearchWorker($QDS$, $index$, $queue[]$, $i$, $N_q$)\;\label{eshq:cw}
		}\label{eshq:ecw}
		Wait for all threads to finish\;\label{eshq:finish}
		\Return ($BSF$)\;\label{eshq:return}
	}
\caption{$Exact Search$}
\label{eshq}
\end{algorithm}

\begin{figure*}[t]
\subfigure[Approximate search for calculating the first BSF\label{fig:MESSIappro}]
{
		\includegraphics[page=1,height=5.6cm]{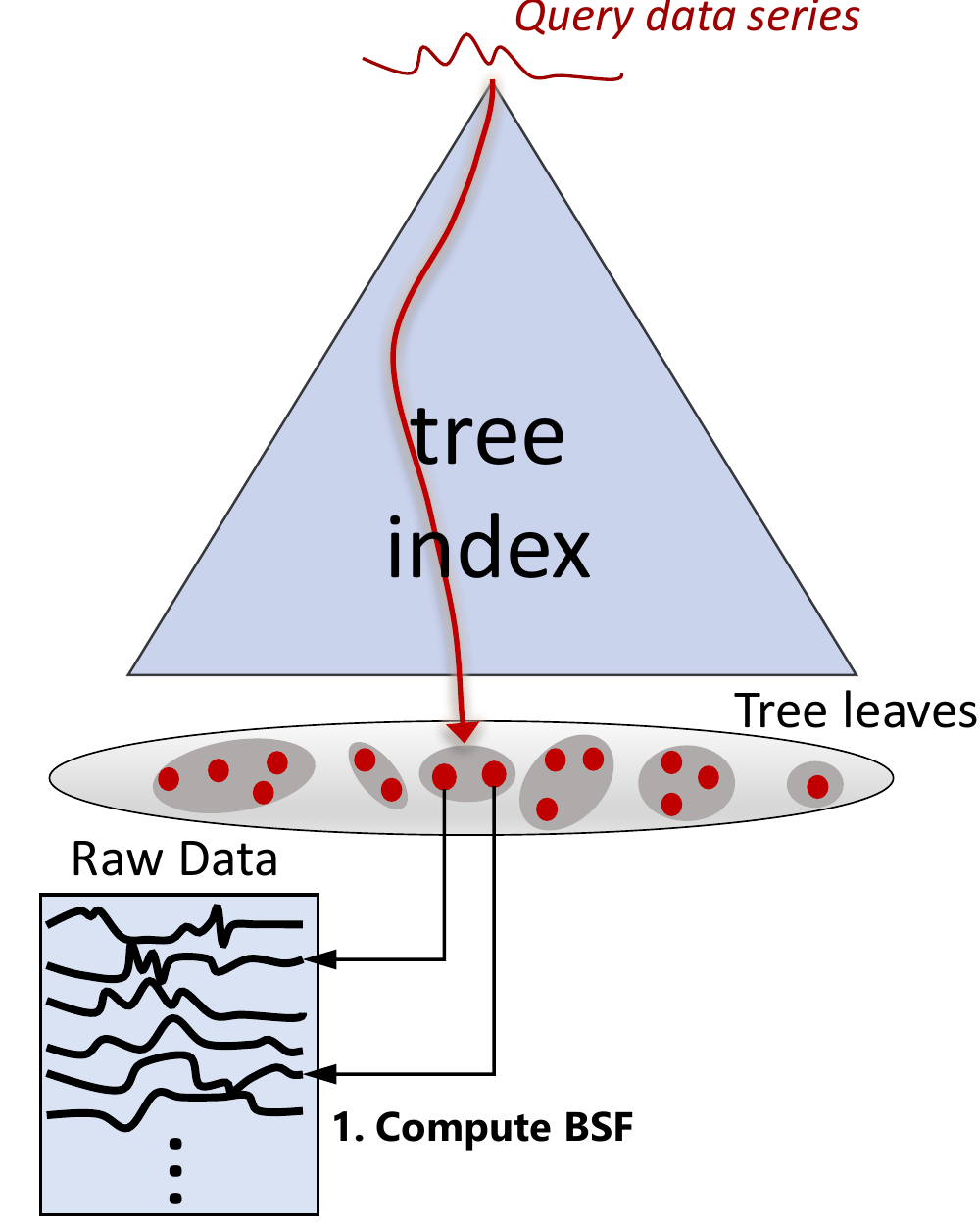}
}
\hspace*{0.3cm}
\subfigure[Tree traversal and node insertion in priority queues\label{fig:MESSIphq}]
{
	\includegraphics[page=1,height=5.6cm]{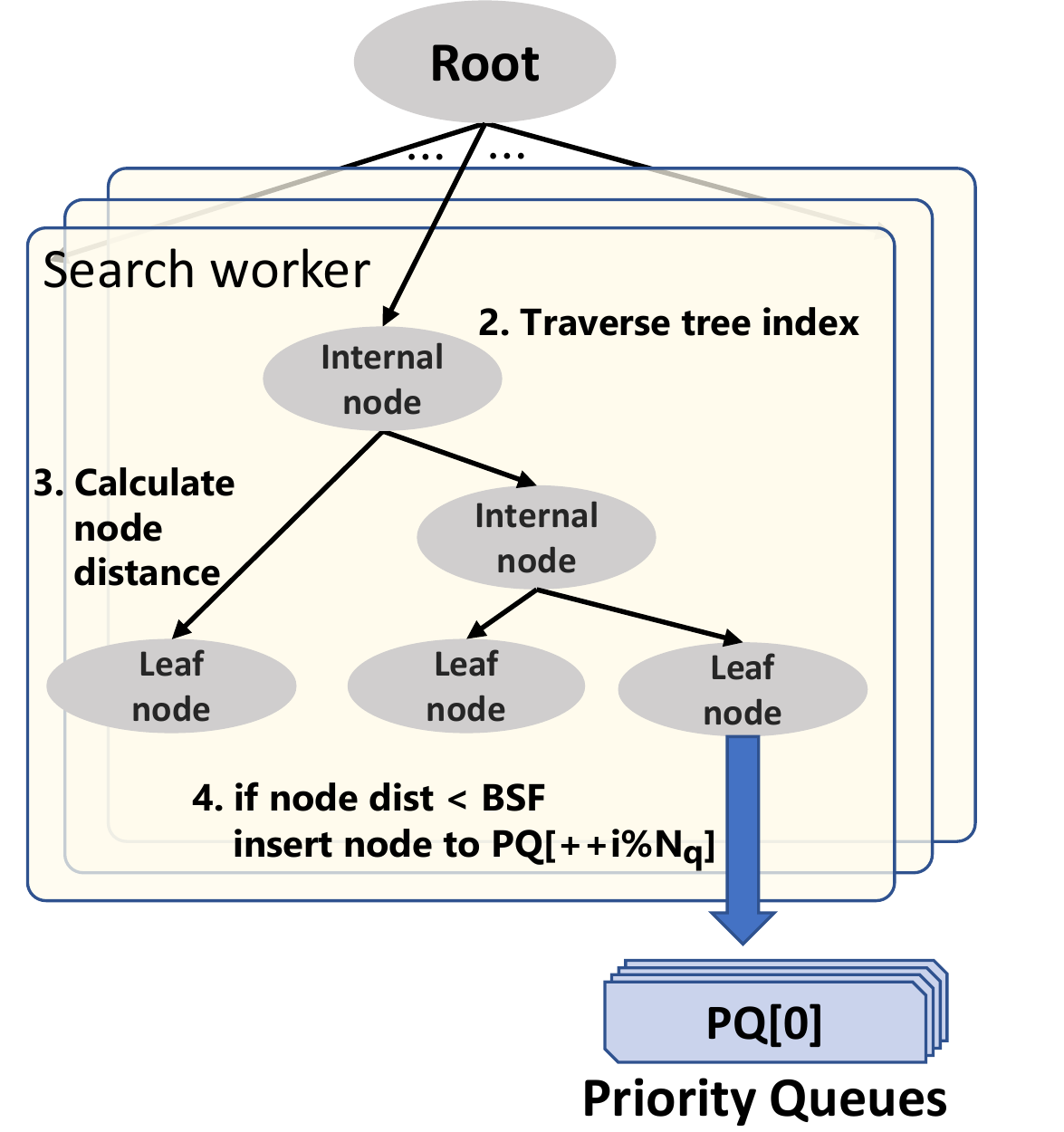}
}
\subfigure[Node distance calculation from priority queues\label{fig:MESSIpophq}]
{
\includegraphics[page=1,height=5.6cm]{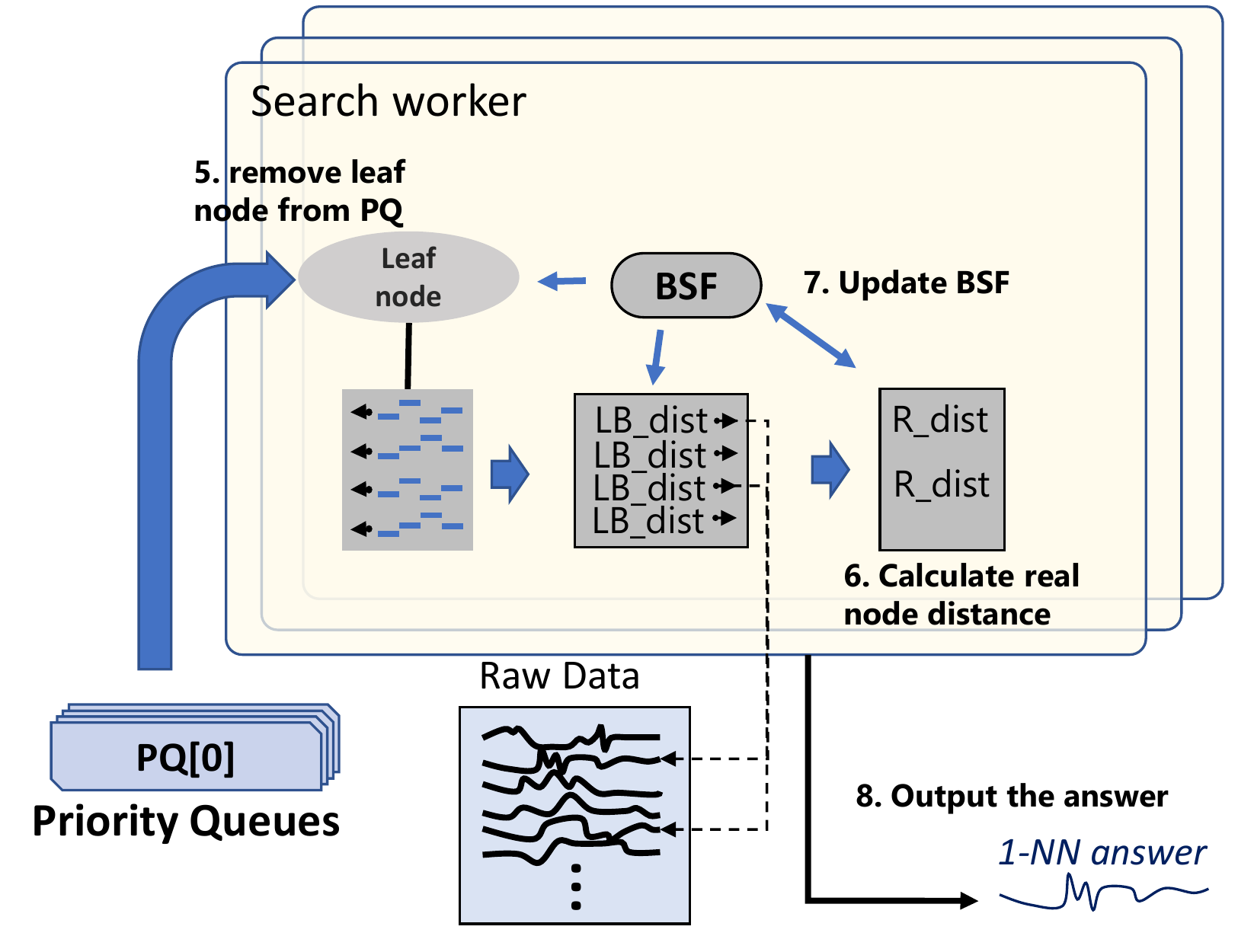}
}
\caption{Workflow and algorithms for MESSI query answering}
\label{fig:MESSIquerychart}
\end{figure*}

The pseudocode for executing an exact search query is shown 
in Algorithm~\ref{eshq}. 
We first calculate the iSAX summary of the query (line~\ref{eshq:calcuqisax}),
and execute an approximate search (line~\ref{eshq:appro}) to find 
the initial value of BSF, i.e., a first upper bound
on the actual distance between the query and the series
indexed by the tree. This process is illustrated in Figure~\ref{fig:MESSIappro}.  

During a search query, the index tree is traversed and the distance of the iSAX summary
of each of the visited nodes to the iSAX summary of the query is calculated. 
If the distance of 
the iSAX summary of a node, $nd$, to the query iSAX summary is higher than BSF, then we are certain that the distances
of all data series indexed by the subtree rooted at $nd$
are higher than BSF. 
So, the entire subtree can be pruned. 
Otherwise, we go down the subtree, and the leaves with a distance to the query smaller 
than the BSF, are inserted in the priority queue.

The technique of using priority queues 
maximizes the pruning degree, thus resulting in a relatively 
small number of raw data series whose real distance to the query series 
must be calculated. As a side effect, BSF converges fast
to the correct value. Thus, the number of iSAX summaries 
that are tested against the iSAX summary of the query series 
is also reduced. 

Algorithm~\ref{eshq} creates $N_s = 48$ threads, called the {\em search workers} (lines~\ref{eshq:scw}-\ref{eshq:ecw}), 
which perform the computation described above by calling $SearchWorker$. 
It also creates $N_q \geq 1$ priority queues (lines~\ref{eshq:scq}-\ref{eshq:ecq}),
where the search workers place those data series that are potential candidates
for real distance calculation. After all search workers have finished (line~\ref{eshq:finish}), 
$ExactSearch$ returns the current value of $BSF$ (line~\ref{eshq:return}). 

We have experimented with two different settings regarding the number of 
priority queues, $N_q$, that the search workers use. 
The first, called {\em Single Queue} ($SQ$), 
refers to $N_q = 1$, whereas the second focuses in the Multiple-Queue ($MQ$) case where $N_q > 1$.
Using a single shared queue imposes a high synchronization overhead, 
whereas using a local queue per thread results in severe load imbalance, since, depending on the workload, the size of the different queues may vary significantly. 
Thus, we choose to use $N_q$ shared queues, 
where $N_q > 1$ is a fixed number
(in our analysis $N_q$ is set to $24$, as experiments our show that this is the
best choice).

The pseudocode of search workers is shown in Algorithm~\ref{eshqw},
and the work they perform is illustrated in Figures~\ref{fig:MESSIphq} and~\ref{fig:MESSIpophq}.
At each point in time, each thread works on a single queue. 
Initially, each queue is shared 
by two threads. 
Each search worker first identifies the queue where it will 
perform its first insertion (line~\ref{eshqw:startq}). 
Then, it repeatedly 
chooses (using Fetch\&Inc)
a root subtree of the index tree to work on 
by calling $TraverseRootSubtree$ (line~\ref{eshqw:traverse}). 
After all root subtrees have been processed (line~\ref{eshqw:barrier}), 
it repeatedly chooses 
a priority queue
(lines~\ref{eshqw:sgetqueue},~\ref{eshqw:egetqueue})
and works on it by calling $ProcessQueue$ (line~\ref{eshqw:getqueue}).  
Each element of the $queue$ array has a field, called $finished$,
which indicates whether the processing of the corresponding priority queue has been
finished. As soon as a search worker determines that all priority queues
have been processed (line~\ref{eshqw:finish}), it terminates.

\begin{algorithm}[tb]
	{\footnotesize
		\SetAlgoLined
		\KwIn{\textbf{QuerySeries} $QDS$, \textbf{Index} $index$, \textbf{Queue} $queue[]$, \textbf{Integer} $pid$, \textbf{Integer} $N_q$}
		\textbf{Shared integer} $N_{b}=0$\;  
		\vspace*{.1cm}
		$q = pid \mod  N_q$\;\label{eshqw:startq}
		\vspace*{.1cm}
		\While{(TRUE)}
		{
			$i\leftarrow${\em Atomically} fetch and increment $N_{b}$\;
			\textbf{if} ($i \geq 2^w$) \textbf{then} break\;
			$TraverseRootSubtree$($QDS$, $index.rootnode[i]$, $queue[]$, $\&q$, $N_q$)\;\label{eshqw:traverse}
			
		}
		\vspace*{.2cm}
		Barrier to synchronize the search workers with one another;\label{eshqw:barrier}\\
                $q = pid \mod N_q$\;
		\vspace*{.2cm}
		
		\While{(true)\label{eshqw:sgetqueue}}
		{
			$ProcessQueue(QDS, index, queue[q])$\label{eshqw:getqueue}\;
			\If{all queue[].finished=true}
			{
			break\;	\label{eshqw:finish}
			}
			$q \leftarrow$ index such that $queue[q]$ has not been processed yet\;
		}\label{eshqw:egetqueue}
	}
\caption{$SearchWorker$}
\label{eshqw}
\end{algorithm}

We continue to describe the pseudocode for 
$TraverseRootSubtree$ 
which is presented in Algorithm~\ref{TraverseRootSubtree} and illustrated in Figure~\ref{fig:MESSIphq}. 
$TraverseRootSubtree$ is recursive. On each internal node, $nd$, it checks whether the (lower bound) 
distance of the iSAX summary of $nd$ to the raw values of the query (line~\ref{insrq:mindist})
is smaller than the current $BSF$, and if it is,
it examines the two subtrees of the node using recursion (lines~\ref{tr:insl}-\ref{tr:insr}). 
If the traversed node is a leaf node and its distance to the 
iSAX summary of the query series is smaller than the current BSF (lines~\ref{tr:sleaf}-\ref{tr:eleaf}), it places it in the appropriate
priority queue (line~\ref{tr:insert}). 
Recall that the priority queues
are accessed in a round-robin fashion (line~\ref{tr:rq}). 
This strategy maintains the size of the queues balanced,
and reduces the synchronization cost of 
node insertions to the queues. 
We implement this strategy by (1) passing a pointer 
to the local variable $q$ of $SearchWorker$
as an argument to $TraverseRootSubtree$, 
(2) using the current value of $q$ for choosing the next queue
to perform an insertion (line~\ref{tr:insert}), and (3) updating the value
of $q$ (line~\ref{tr:rq}). 
Each queue
may be accessed by more than one threads, so a lock per queue is
used to protect its concurrent access by multiple threads.

\begin{algorithm}[tb]
	{\footnotesize
		\SetAlgoLined
		\KwIn{\textbf{QuerySeries} $QDS$, \textbf{Node} $node$, \textbf{queue} $queue[]$, \textbf{Integer} $*pq$, \textbf{Integer} $N_q$}
		\vspace*{.2cm}
		$nodedist$ = FindDist($QDS$, $node$)\;\label{insrq:mindist}
		\uIf{$nodedist$ $>$ $BSF$} 
		{
			break\;
		}
		\uElseIf{$node$ is a leaf\label{tr:sleaf}} 
		{
			acquire $queue[*pq]$ lock\;
			Put $node$ in $queue[*pq]$ with priority $nodedist$\;\label{tr:insert}
			release $queue[*pq]$ lock\;
			\emph{// next time, insert in the subsequent queue\\}
			$*pq\leftarrow (*pq+1) \mod N_q$\;\label{tr:rq}
		}\label{tr:eleaf}
		\Else
		{
			TraverseRootSubtree($node.leftChild,queue[],pq,N_q$)\;\label{tr:insl}
			TraverseRootSubtree($node.rightChild,queue[],pq,N_q$)\label{tr:insr}
		}
	}
	\caption{$TraverseRootSubtree$}
	\label{TraverseRootSubtree}
\end{algorithm}

We next describe how $ProcessQueue$ works (see Algorithm~\ref{ProcessQueue} and Figure~\ref{fig:MESSIpophq}). 
The search worker repeatedly 
removes the (leaf) node, $nd$, with the highest priority 
from the priority queue, 
and checks whether the corresponding distance stored in the queue 
is still less than the BSF. 
We do so, because the
BSF may have changed since the time that the leaf node was inserted in the priority queue. 
If the distance is less than the BSF, then $CalculateRealDistance$ (line~\ref{process:rd1}) is called, in order to identify if any series in the leaf node (pointed to by $nd$)
has a real distance to the query that is smaller than the current BSF. 
If we discover such a series (line~\ref{process:if}), 
$BSF$ is updated to the new value (line~\ref{process:upbsf}). 
We use a lock to protect BSF from concurrent update efforts (lines~\ref{process:loc},~\ref{process:unloc}).
Previous experiments showed that the initial value of BSF is very close to its final value~\cite{gogolou2019progressive}. 
Indeed, in our experiments, the BSF is updated only
	10-12 times (on average) per query. 
	So, 
	the synchronization cost for updating the BSF is negligible. 

\begin{algorithm}[tb]
	{\footnotesize	
		\SetAlgoLined
		\KwIn{\textbf{QuerySeries} $QDS$, \textbf{Index} $index$, \textbf{Queue} $Q$}
		\While{$node$ = DeleteMin($Q$)\label{process:popnode}}
		{
			\uIf{$node.dist$ $<$ $BSF$}
			{
				$realDist$ = CalculateRealDistance($QDS$, $index$, $node$)\label{process:rd1}\;
				\If{\textbf{$realDist$} $<$ $BSF$\label{process:if}} 
				{
					acquire $BSFLock$\;\label{process:loc}
					$BSF$ = $realDist$\;\label{process:upbsf}
					release $BSFLock$\;\label{process:unloc}
				}
			}
			\Else
			{
				$q.finished$ = true\;
				break\;	
			}	
		}
	}
\caption{$ProcessQueue$}
\label{ProcessQueue}
\end{algorithm}

In Algorithm~\ref{realdist}, we depict the pseudocode for $CalculateRealDistance$. 
Note that we perform the real distance calculation using SIMD. 
However, the use of SIMD does not have the same significant
impact in performance as in ParIS~\cite{peng2018paris}. 
This is because pruning is much more
effective in MESSI, since for each candidate series in the examined leaf node, 
$CalculateRealDistance$ first performs a lower bound distance calculation,
and proceeds to the real distance calculation only if necessary (line~\ref{rd:rd}). 
Therefore, the number
of (raw) data series to be examined is limited in comparison
to those examined in ParIS (we quantify the effect of this new design in our experimental evaluation).

\begin{algorithm}[tb]
{\footnotesize
	\SetAlgoLined
	\KwIn{\textbf{QuerySeries} $QDS$, \textbf{Index} $index$, \textbf{node} $node$, \textbf{float} $BSF$}

	\For{every ($isax$, $pos$) pair $\in$ $node$} 
	{
		\If{$LowerBound\_SIMD$($QDS$, $isax$) $<$ $BSF$\label{rd:ld}}
		{
			$dist = RealDist\_SIMD(index.RawData[pos],QDS)$\;\label{rd:rd}
			\If{$dist < BSF$}
			{
				 $BSF = dist$\;
			}
		}
	}
	\Return{($BSF$)}
}
\caption{$CalculateRealDistance$}
\label{realdist}
\end{algorithm}


%% file: experiments.tex
\section{Experimental Evaluation}
\label{sec:experiments}

In this section, we present our experimental evaluation. We use synthetic 
and real datasets in order to compare the performance of MESSI with that of competitors 
that have been proposed in the literature and baselines that we developed. 
We demonstrate that, under the same settings, MESSI is able to construct the index up 
to 4.2x faster, and answer similarity search queries up to 11.2x faster than the competitors. 
Overall, MESSI exhibits a robust performance across different datasets and settings, and 
enables for the first time the exploration of very large data series collections 
at interactive speeds. 

\subsection{Setup}
\label{setupsec}
We used a server with 2x Intel Xeon E5-2650 v4 2.2Ghz CPUs 
(12 cores/24 hyper-threads each) and 256GB RAM.
All algorithms were implemented in C, and compiled using GCC v6.2.0 on Ubuntu Linux v16.04.

\noindent\textbf{[Algorithms]} 
We compared MESSI to the following algorithms: (i) ParIS~\cite{peng2018paris}, 
the state-of-the-art modern hardware data series index.
(ii) ParIS-TS, our extension of ParIS, where we implemented in a parallel fashion the traditional tree-based 
exact search algorithm~\cite{shieh2008sax}.
In brief, this algorithm traverses the tree, and concurrently (1) inserts 
in the priority queue the nodes (inner nodes or leaves) that cannot be pruned based on the lower bound distance,
and (2) pops from the queues nodes for which it calculates the real distances to the candidate series~\cite{shieh2008sax}. 
In contrast, MESSI (a) first makes a \emph{complete pass} over the index using lower bound distance 
computations and then proceeds with the real distance computations; 
(b) it only considers the \emph{leaves} of the index for insertion in the priority queue(s); and (c) performs a \emph{second} filtering step 
using the lower bound distances when popping elements from the priority queue (and before computing 
the real distances). The performance results we present later justify the choices we have made in MESSI, 
and demonstrate that a straight-forward implementation of tree-based exact search leads to sub-optimal 
performance. 
(iii) UCR Suite-P, our parallel implementation of 
the state-of-the-art optimized serial scan technique, UCR Suite~\cite{rakthanmanon2012searching}. 
In UCR Suite-P, every thread is assigned a part of the in-memory data series array, and all threads  
concurrently and independently process their own parts, performing the real distance calculations in SIMD, 
and only synchronize at the end to produce the final result.
(We do not consider the non-parallel UCR Suite version in our experiments, since it is almost 300x slower.)
All algorithms operated exclusively in main memory (the datasets were already loaded in memory, as well).
The code for all algorithms used in this paper is available online~\cite{sourcescode}.

\noindent\textbf{[Datasets]} 
In order to evaluate the performance of the proposed approach, 
we use several synthetic datasets for a fine grained analysis, 
and two real datasets from diverse domains.
Unless otherwise noted, the series have a size of 256 points, 
which is a standard length used in the literature, 
and allows us to compare our results to previous work.
We used synthetic datasets of sizes 50GB-200GB (with a default size of 100GB),
and a random walk data series generator that works as follows: 
a random number is first drawn from a Gaussian distribution N(0,1), 
and then at each time point a new number is drawn from this distribution 
and added to the value of the last number. 
This kind of data generation has been extensively used in the past 
(and has been shown to model real-world financial data)~\cite{yi2000fast,shieh2008sax,wang2013data,isax2plus,zoumpatianos2016ads}.
We used the same process to generate 100 query series. 


For our first real dataset, \emph{Seismic}, 
we used the IRIS Seismic Data Access repository~\cite{iris} to gather 100M series representing seismic waves 
from various locations,  
for a total size of 100GB.
The second real dataset, \emph{SALD}, includes neuroscience MRI data series~\cite{url:SALD}, 
for a total of 200M series of size 128, of size 100 GB.
In both cases, we used as queries 100 series out of the datasets (chosen using our synthetic series generator).

In all cases, we repeated the experiments 10 times and we report the average values. 
We omit reporting the error bars, since all runs gave results that were very similar (less than 3\% difference).
Queries were always run in a sequential fashion, one after the other, 
in order to simulate an exploratory analysis scenario, 
where users formulate new queries after having seen the results of the previous one.


\begin{figure*}[tb]
	\begin{minipage}[b]{0.23\textwidth}
		
		\includegraphics[page=1,width=\columnwidth]{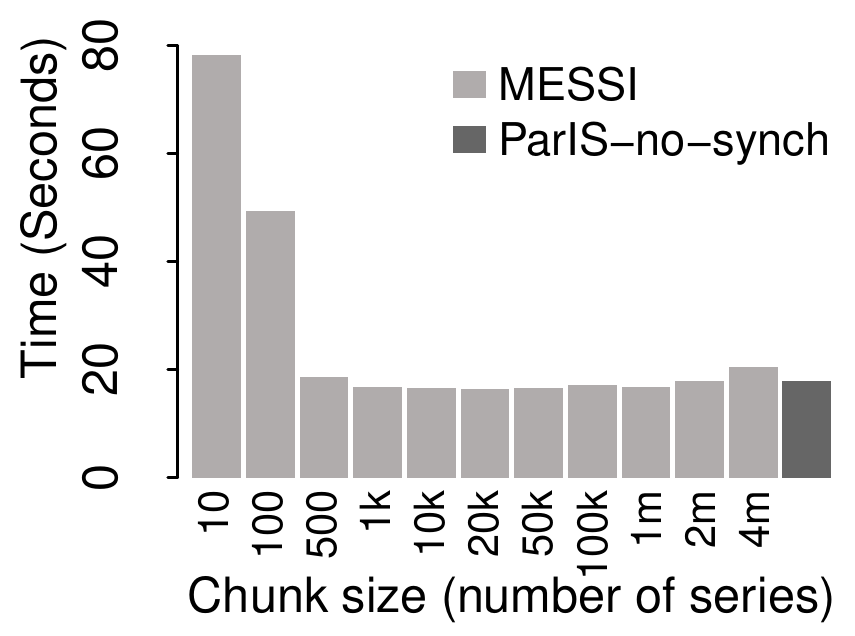}
		\caption{{Index creation, vs. chunk size}}
		\label{fig:chunksize}
	\end{minipage}
	\begin{minipage}[b]{0.225\textwidth}
		
		\includegraphics[page=1,width=\columnwidth]{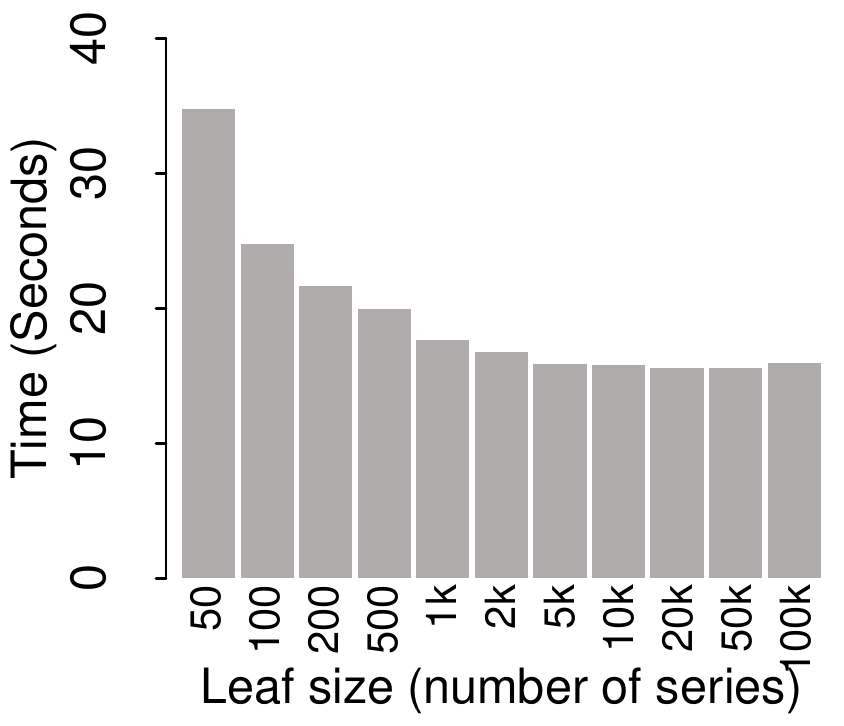}
		\caption{Index creation, vs. leaf size}
		\label{fig:indleaf}
	\end{minipage}
	\begin{minipage}[b]{0.31\textwidth}
		
		\includegraphics[page=1,width=\columnwidth]{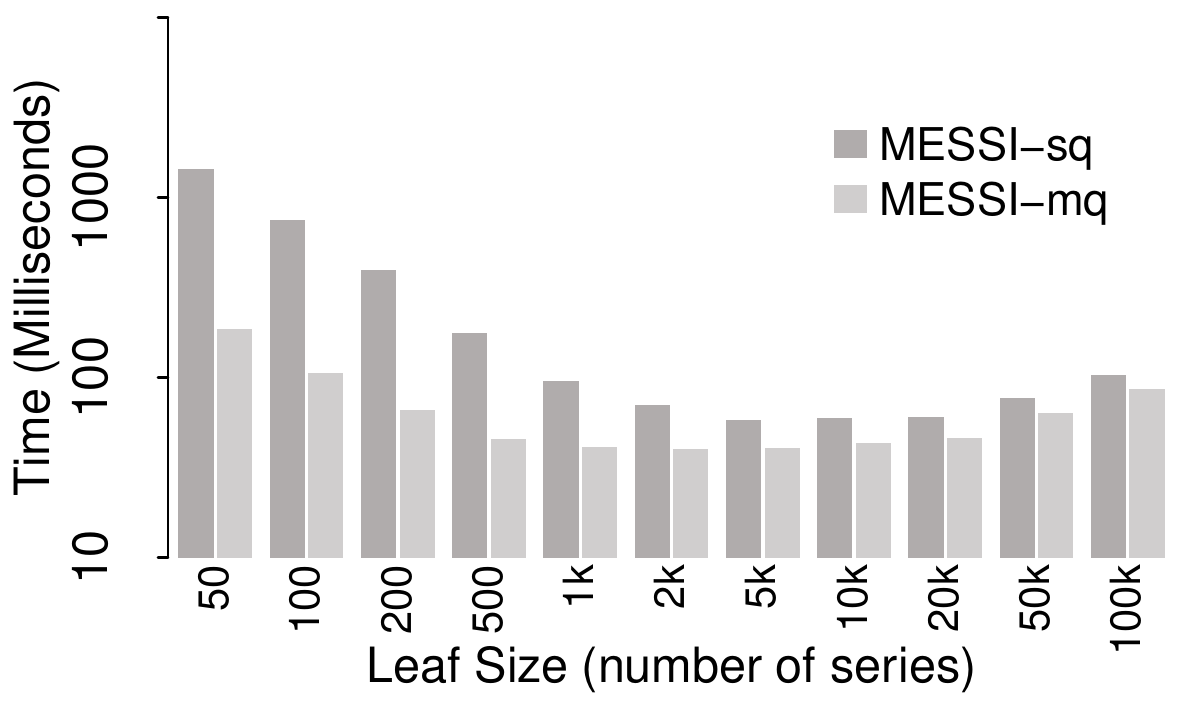}
		\caption{Query answering, vs. leaf size}
		\label{fig:dleaf}
	\end{minipage}
	\begin{minipage}[b]{0.22\textwidth}
		
		\includegraphics[page=1,width=\columnwidth]{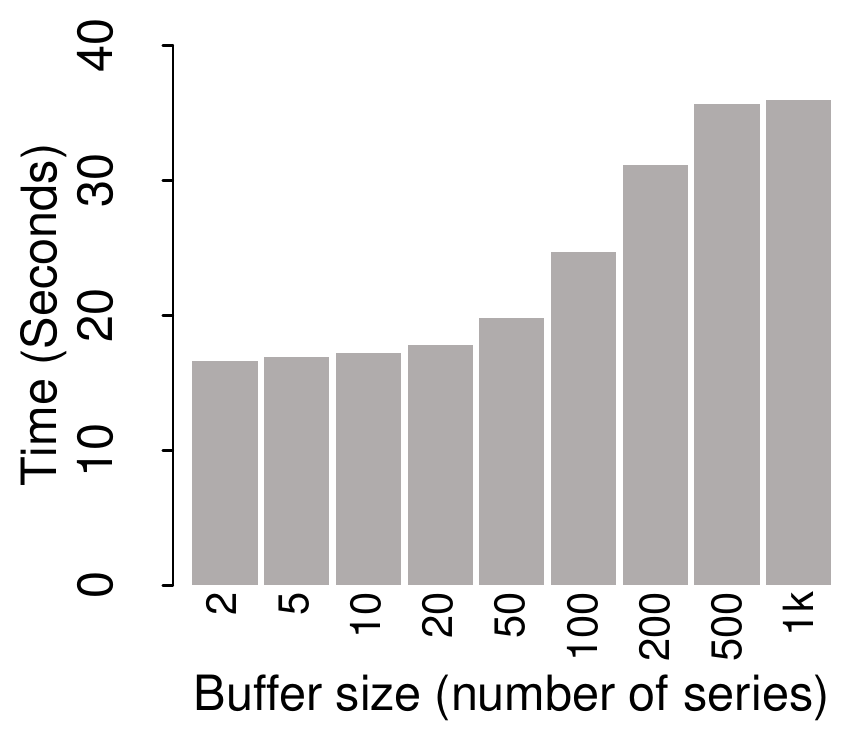}
		\caption{Index creation, vs. initial iSAX buffer size}
		\label{fig:initialisaxbuffersize}
	\end{minipage}
\end{figure*}

\subsection{Parameter Tuning Evaluation}
\label{parameter}
In all our experiments, we use 24 index workers and 48 search workers.
We have chosen the chunk size to be 20MB (corresponding to 20K series of length 256 points).
Each part of any iSAX buffer, initially holds a small constant number of data series, 
but its size changes dynamically depending on how many data
series it needs to store. 
The capacity of each leaf of the index tree
is 2000 data series (2MB). 
For query answering, MESSI-mq utilizes 24 priority queues (whereas MESSI-sq
utilizes just one priority queue). 
In either case, each priority queue is implemented 
using an array whose size changes dynamically based on how many elements must be stored in it.
Below we present the experiments that justify the choices for these parameters.

\begin{figure}[tb]
	\centering
	\includegraphics[page=1,width=0.9\columnwidth]{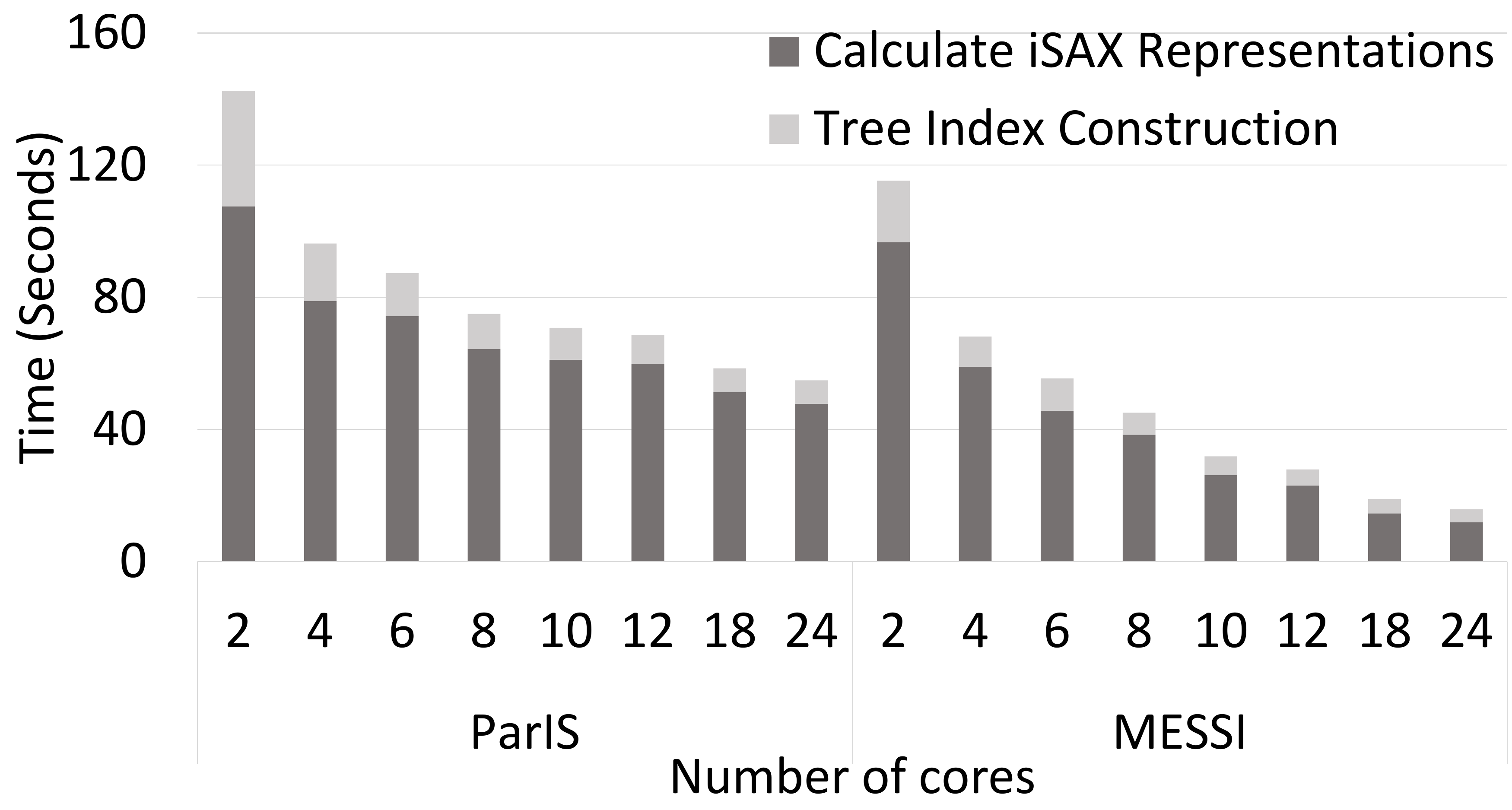}
	\caption{Index creation, varying number of cores}
	\label{fig:pRecBuf}
\end{figure}

\begin{figure*}[tb]
	
	\begin{minipage}[b]{0.24\textwidth}
		\centering
		\includegraphics[page=1,width=\columnwidth]{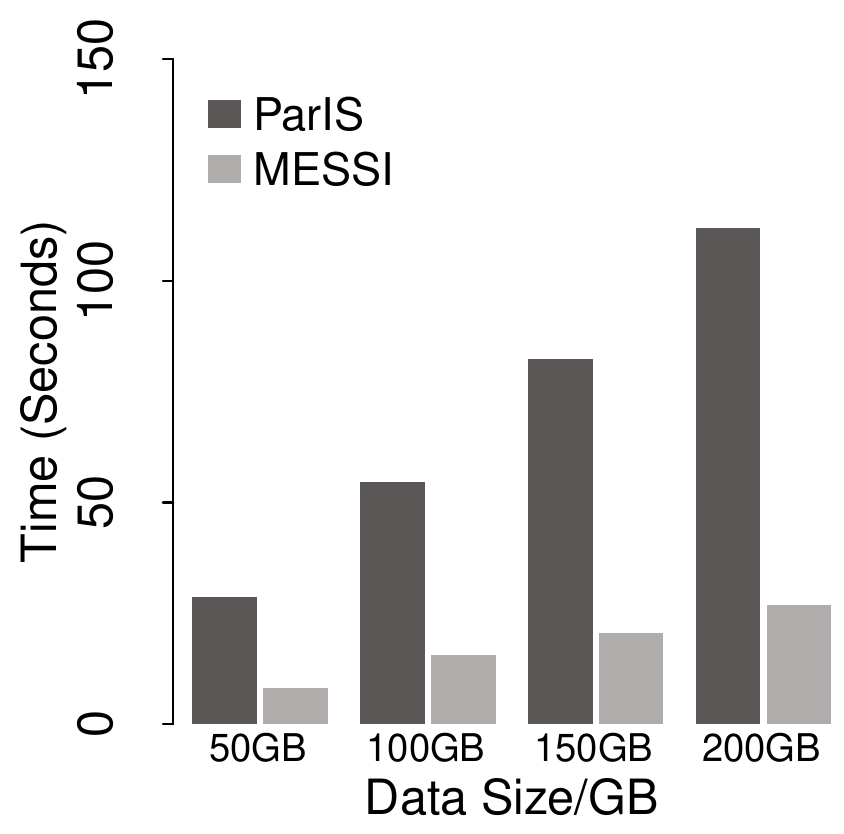}
		\caption{Index creation, vs. data size}
		\label{fig:incvarysize}
	\end{minipage}
	\begin{minipage}[b]{0.38\textwidth}
		\centering
		\includegraphics[page=1,width=\columnwidth]{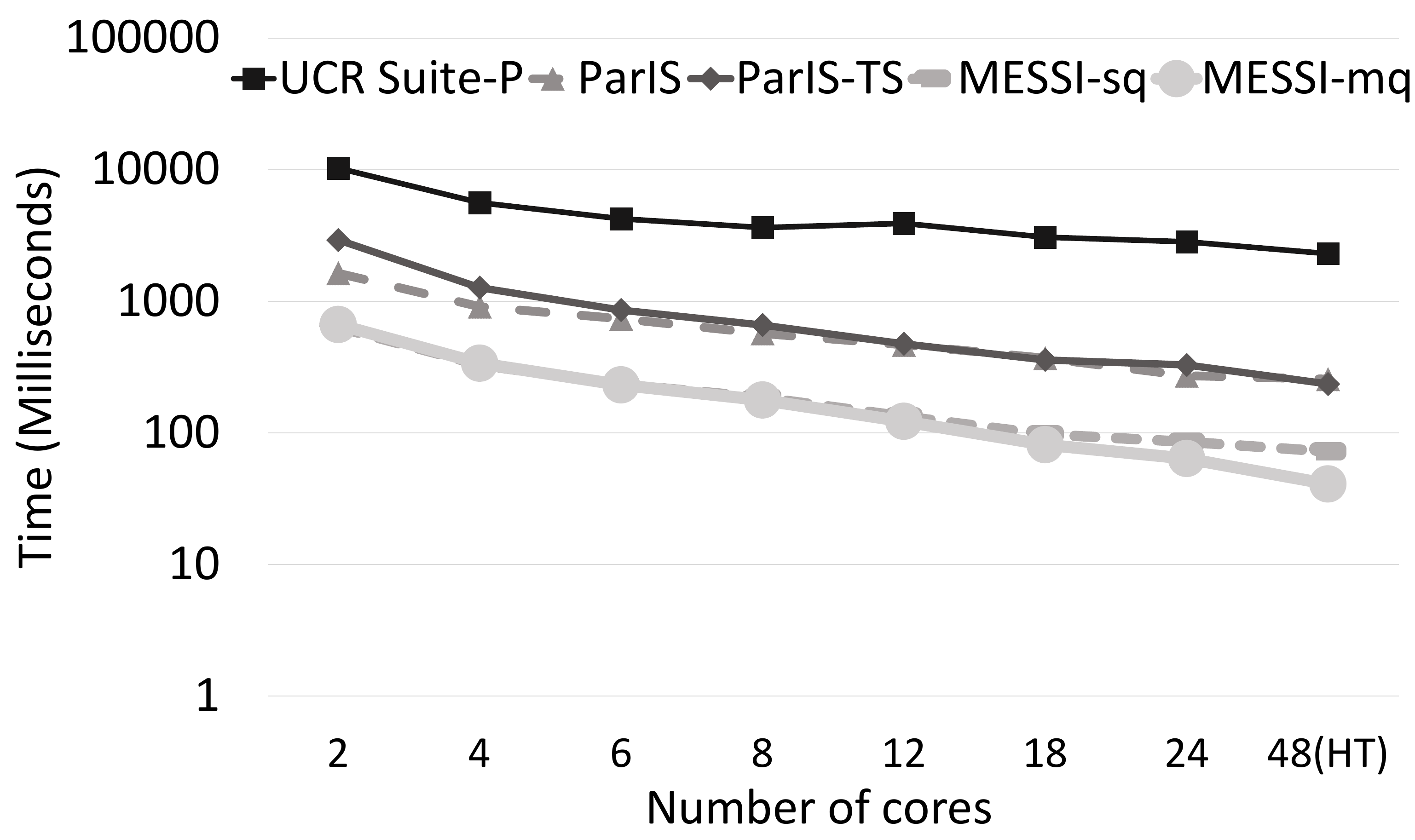}
		\caption{Query answering, vs. number of cores}
		\label{fig:varycore}
	\end{minipage}
	\begin{minipage}[b]{0.38\textwidth}
		\centering
		\includegraphics[page=1,width=\columnwidth]{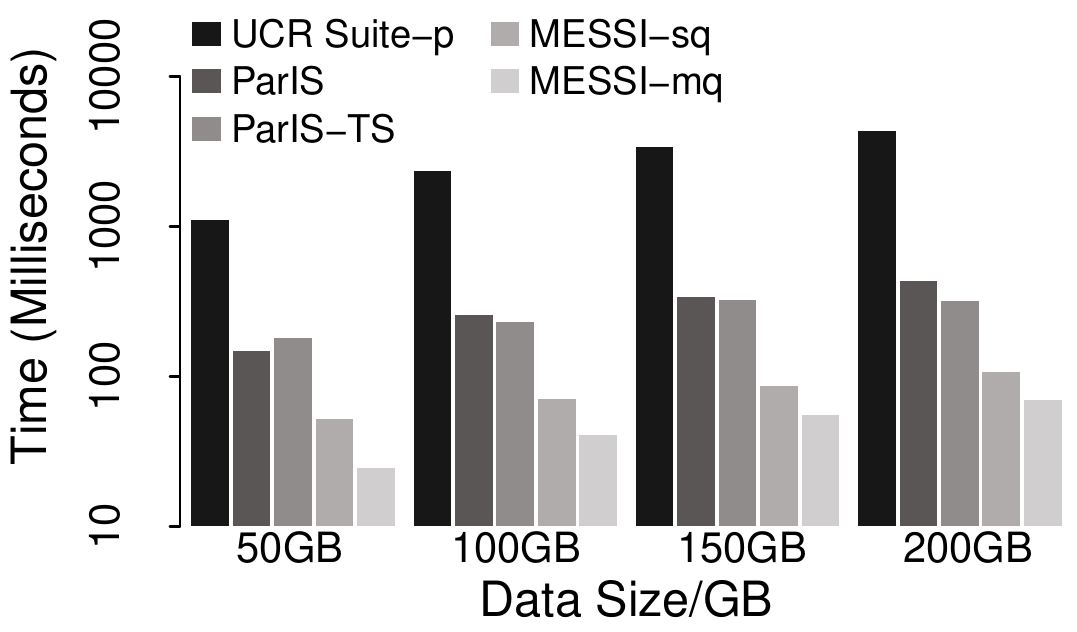}
		\caption{Query answering, vs. data size}
		\label{fig:scalquery}
	\end{minipage}
\end{figure*}

Figure~\ref{fig:chunksize} illustrates the time it takes MESSI to build the tree index for different chunk sizes on a random dataset of 100GB. 
The required time to build the index decreases when the chunk size is small and does not have any big influence in performance after the value of 1K (data series). Smaller chunk sizes than 1K result in high contention when accessing the fetch\&increment object used to assign chunks to index workers.
In our experiments, we have chosen a size of 20K, as this gives slightly better performance than setting it to 1K.      

Figures~\ref{fig:indleaf} and~\ref{fig:dleaf} show the impact that varying the leaf size of the tree index has in the time needed for the index creation and for query answering, respectively. As we see in Figure~\ref{fig:indleaf}, 
the larger the leaf size is, the faster index creation becomes.
However, once the leaf size becomes 5K or more, this time improvement
is insignificant. On the other hand, Figure~\ref{fig:dleaf} shows that the query answering time takes its minimum value when the leaf size is set to 2K (data series). 
So, we have chosen this value for our experiments.  

Figure~\ref{fig:dleaf} indicates that the influence of varying the leaf size is significant for query answering. Note that when the leaf size is small, there are more leaf nodes in the index tree and therefore, it is highly probable that more nodes will be inserted in the queues, and vice versa. 
On the other hand, as the leaf size increases, the number of real distance calculations that are performed to process each one of the leaves in the queue is larger. 
This causes
load imbalance among the different search workers that process the priority queues.    
For these reasons, we see that at the beginning the time goes down as the leaf size increases, it reaches its minimum value for leaf size 2K series, and then it goes up again as the leaf size further increases. 

Figure~\ref{fig:initialisaxbuffersize} shows the influence of the initial iSAX buffer size during index creation. This initialization cost is not negligible given that we allocate $2^w$ iSAX buffers, each consisting of $24$ parts (recall that 24 is the number of index workers in the system).  
As expected, the figure illustrates that smaller initial sizes for the buffers result in better performance. 
We have chosen the initial size of each part of the iSAX buffers to be a small constant number of data series. 
(We also considered an alternative design that collects statistics and allocates the iSAX buffers right from the beginning, but was slower.)

\begin{figure}[tb]
	\centering
	\includegraphics[page=1,width=0.9\columnwidth]{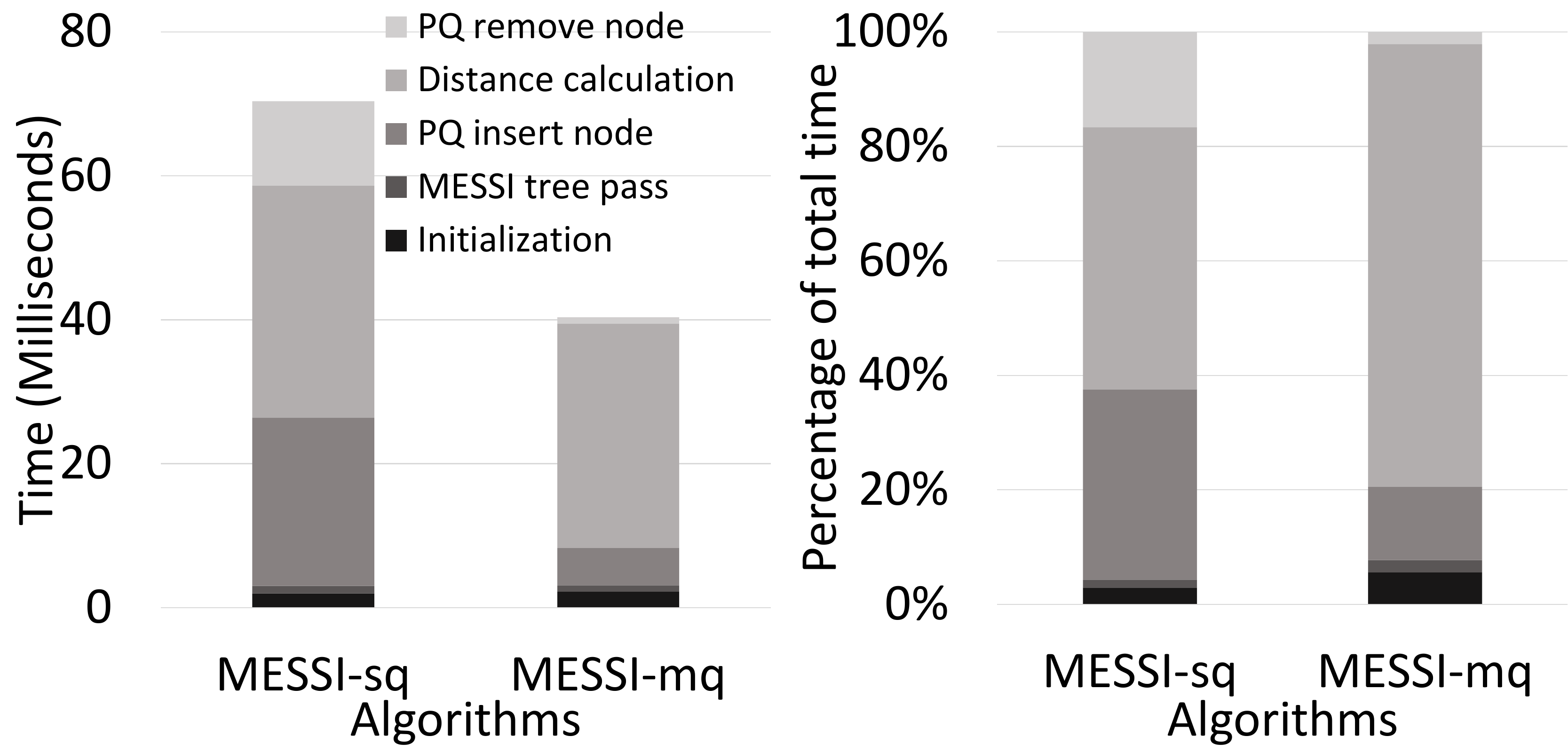}
	\caption{Query answering with different queue type}
	\label{fig:diffq}
\end{figure}

\begin{figure}[tb]
	\centering
	\includegraphics[page=1,width=0.9\columnwidth]{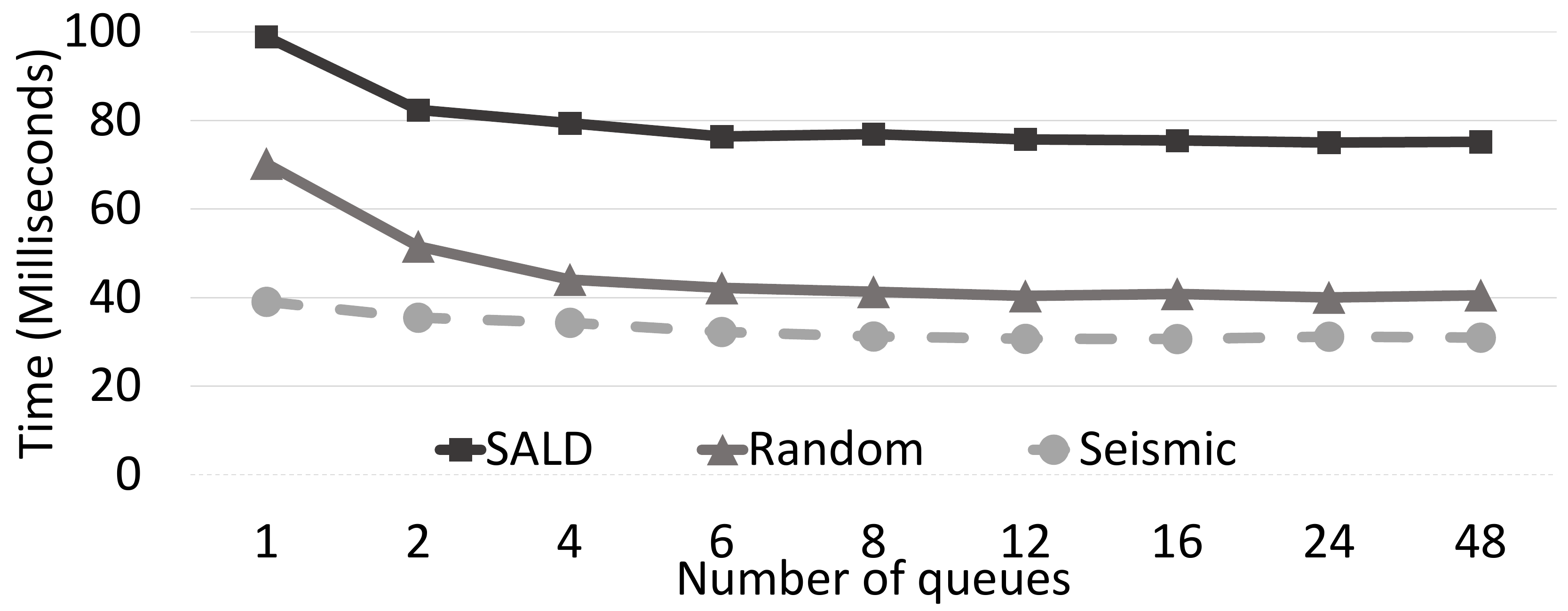}
	\caption{Query answering, vs. number of queues}
	\label{fig:dqueue}
\end{figure}

We finally justify the choice of using more than one priority queues
for query answering. 
%
As Figure~\ref{fig:varycore} shows, MESSI-mq and MESSI-sq have similar performance when the number of threads is smaller than 24. 
However, as we go from 24 to 48 cores, the synchronization cost for accessing the single priority queue in MESSI-sq has negative impact in performance.
Figure~\ref{fig:diffq}  presents the breakdown of the query answering time for these two algorithms. 
The figure shows that in MESSI-mq, the time needed to insert and remove nodes from the list is significantly reduced. 
As expected, the time needed for the real distance calculations and for the tree traversal are about the same in both algorithms. 
This has the effect that the time needed for the distance calculations becomes the dominant factor. 
The figure also illustrates the percentage of time that goes on each of these tasks.    
Finally, Figure~\ref{fig:dqueue} illustrates the impact that the number of priority queues has in query answering performance.
As the number of priority queues increases, the time goes down,
and it takes its minimum value when this number becomes 24. 
So,
we have chosen this value for our experiments.


\subsection{Comparison to Competitors}
\noindent{\bf [Index Creation]}
Figure~\ref{fig:pRecBuf} compares the index creation time of MESSI with  that 
of ParIS as the number of cores increases for a dataset of 100GB. 
The time MESSI needs for index creation 
is significantly smaller than that of ParIS. 
Specifically, MESSI is  3.5x faster than ParIS. 
The main reasons for this are on the one hand 
that MESSI exhibits lower contention cost when accessing the iSAX buffers
in comparison to the corresponding cost paid by ParIS, 
and on the other hand, that MESSI achieves better load balancing when performing the computation
of the iSAX summaries from the raw data series.
Note that due to synchronization cost, 
the performance improvement that both algorithms exhibit decreases 
as the number of cores increases; this trend is more prominent in ParIS, while MESSI manages to exploit to a larger degree the available hardware.

In Figure~\ref{fig:incvarysize}, we depict the index creation time as the dataset size 
grows from 50GB to 200GB. 
We observe that MESSI performs up to 4.2x faster than ParIS (for the 200GB dataset), 
with the improvement becoming larger with the dataset size.


\begin{figure}[tb]
	\centering
\begin{minipage}[b]{0.3\columnwidth}
	\includegraphics[page=1,width=\columnwidth]{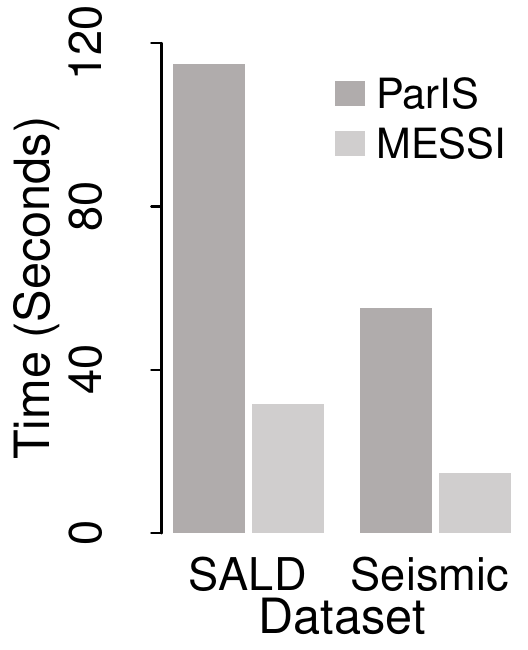}
	\caption{Index creation for real datasets}
	\label{fig:realinc}
\end{minipage}
\hspace*{0.2cm}
\begin{minipage}[b]{0.5\columnwidth}
	\includegraphics[page=1,width=\columnwidth]{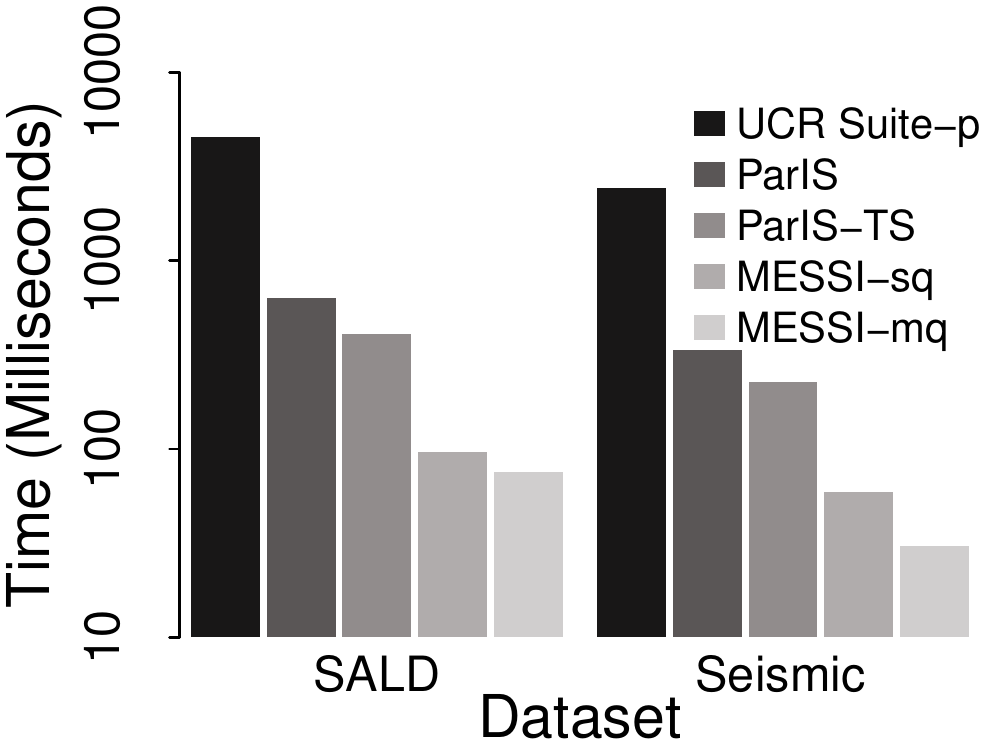}
	\caption{Query answering for real datasets}
	\label{fig:realqa}
\end{minipage}
\end{figure}
\begin{figure}[tb]
	\centering
\begin{minipage}[b]{0.95\columnwidth}
	\subfigure[Lower bound distance calculations
	\label{fig:ndist1}]{
		\includegraphics[page=1,width=0.45\columnwidth]{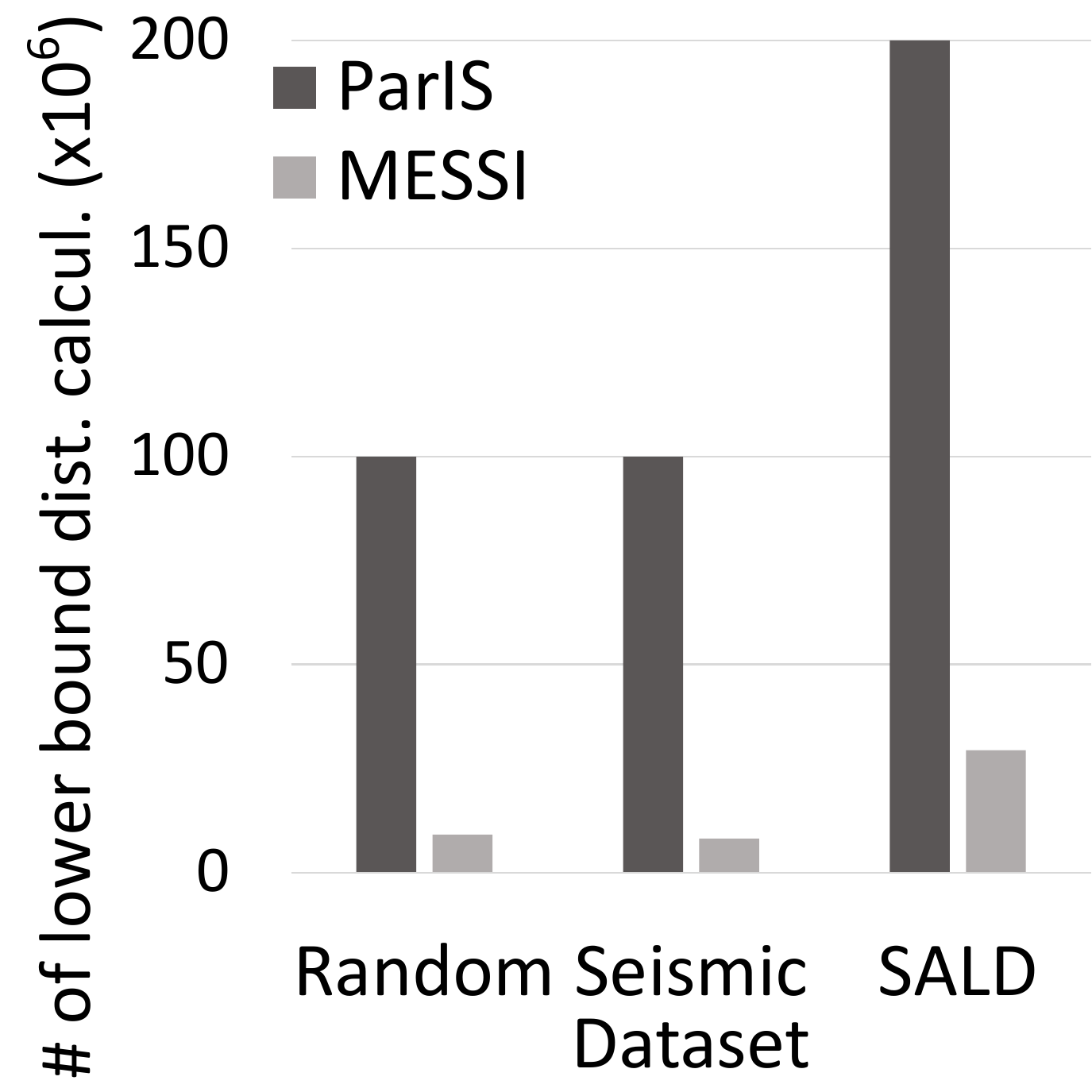}
	}
	\subfigure[Real distance calculations 
	\label{fig:ndist2}]{
		\includegraphics[page=2,width=0.45\columnwidth]{picture/calculationnumber2}
	}
	\caption{Number of distance calculations}
	\label{fig:ndist}
\end{minipage}

\end{figure}



\noindent{\bf [Query Answering]}
%
Figure~\ref{fig:varycore} 
compares the performance of the MESSI query answering algorithm
to its competitors, as the number of cores increases, for a random dataset of 100GB
(y-axis in log scale). 
The results show that both MESSI-sq and MESSI-mq perform much better
than all the other algorithms. 
Note that the performance of MESSI-mq is better than that of MESSI-sq,
so when we mention MESSI in our comparison below we refer to MESSI-mq. 
MESSI is 55x faster than UCR Suite-P and 6.35x faster than ParIS 
when we use 48 threads (with hyperthreading). In contrast to ParIS, 
MESSI applies pruning when performing the lower bound distance calculations
and therefore it executes this phase much faster. 
Moreover, the use of the priority queues result in 
even higher pruning power. As a side effect, MESSI also performs 
less real distance calculations than ParIS. 
Note that UCR Suite-P does not perform any pruning, thus resulting 
in a much lower performance than the other algorithms. 

Figure~\ref{fig:scalquery} shows that
this superior performance of MESSI is exhibited  
for different data set sizes as well.
Specifically, MESSI is up to 61x faster than 
UCR Suite-p (for 200GB), up to 6.35x faster than ParIS (for 100GB), 
and up to 7.4x faster than ParIS-TS (for 50GB).

\noindent{
{\bf [Performance Benefit Breakdown]}
Given the above results, we now evaluate several of the design choices of MESSI in isolation. 
Note that some of our design decisions stem from the fact that in our index 
the root node has a large number of children. 
Thus, the same design ideas are applicable to the iSAX family of indices~\cite{lernaeanhydra} (e.g., iSAX2+, ADS+, ULISSE).
Other indices however~\cite{lernaeanhydra}, use a binary tree (e.g., DSTree), or a tree with a very small fanout (e.g., SFA trie, M-tree), so new design techniques are required for efficient parallelization. 
However, some of our techniques, e.g., 
the use of (more than one) priority queue, the use of SIMD,
and some of the data structures designed to reduce the syncrhonization cost 
can be applied to all other indices.
Figure~\ref{fig:stepbystepquery} shows the results for the query answering performance. 
The leftmost bar (ParIS-SISD) shows the performance 
of ParIS when SIMD is \emph{not} used. 
By employing SIMD,
ParIS becomes 60\% faster than ParIS-SISD. 
We then measure the performance for ParIS-TS, which is about 10\% faster than ParIS.
This performance improvement comes form the fact that using 
the index tree (instead of the SAX array that ParIS uses) to prune the search space and determine the data series for which 
a real distance calculation must be performed, significantly reduces the number of lower bound distance 
calculations. 
ParIS calculates lower bound distances for all the 
data series in the collection, and pruning is performed only when calculating
real distances, whereas in ParIS-TS pruning occurs when calculating lower bound
distances as well. 

MESSI-mq further improves performance by only inserting in the priority queue leaf nodes (thus, reducing the size of the queue), and by using multiple  queues (thus, reducing the synchronization cost). 
This makes MESSI-mq 83\% faster than ParIS-TS.

\noindent{\bf [Real Datasets]}
Figures~\ref{fig:realinc} and~\ref{fig:realqa}  reaffirm that MESSI
exhibits the best performance for both index creation and query answering, 
even when executing on the real datasets, SALD and Seismic (for a 100GB dataset). The reasons for this
are those explained in the previous paragraphs. 
Regarding index creation, MESSI is 3.6x faster than ParIS on SALD and 3.7x faster than ParIS 
on Seismic, for a 100GB dataset. 
Moreover, for SALD, MESSI query answering is 60x faster than UCR Suite-P 
and 8.4x faster than ParIS, whereas for Seismic,
it is 80x faster than UCR Suite-P, and almost 11x faster than ParIS.
Note that MESSI exhibits better performance than UCR Suite-P 
in the case of real datasets. This is so because working on random
data results in better pruning than that on real data. 

Figures~\ref{fig:ndist1} and~\ref{fig:ndist2} illustrate the number of lower bound and real distance calculations,
respectively, performed by the different query algorithms on the three datasets. 
ParIS calculates the distance between the iSAX summaries of every single data series 
and the query series (because, as we discussed in Section~\ref{sec:prelim}, it implements the SIMS strategy for query answering). 
In contrast, MESSI performs pruning even during the lower bound distance calculations, 
resulting in much less time for executing this computation. 
Moreover, this results in a significantly reduced number of data series whose 
real distance to the query series must be calculated. 

The use of the priority queues lead to even less real distance calculations, 
because they help the BSF to converge faster to its final value. 
MESSI performs no more than 15\% of the lower bound distance calculations
performed by ParIS.

\noindent{\bf [MESSI with DTW]}
In our final experiments, we demonstrate that MESSI not only accelerates similarity search based on Euclidean distance, but can also be used to significantly accelerate similarity search using the Dynamic Time Warping (DTW) distance measure~\cite{berndt1994using}.
We note that no changes are required in the index structure; we just have to build the envelope of the LB\_Keogh method~\cite{keogh2005exact} around the query series, and then search the index using this envelope.
%
%
%
Figure~\ref{fig:dtw} shows the query answering time for different dataset sizes (we use a warping window size of 10\% of the query series length, which is commonly used in practice~\cite{keogh2005exact}). 
The results show that MESSI-DTW is up to 34x faster than UCR Suite-p DTW (and more than 3 orders of magnitude faster than the non-paralell version of UCR Suite DTW).

\begin{figure}[tb]
		\begin{minipage}[b]{0.4\columnwidth}
		\centering
		\includegraphics[width=\columnwidth]{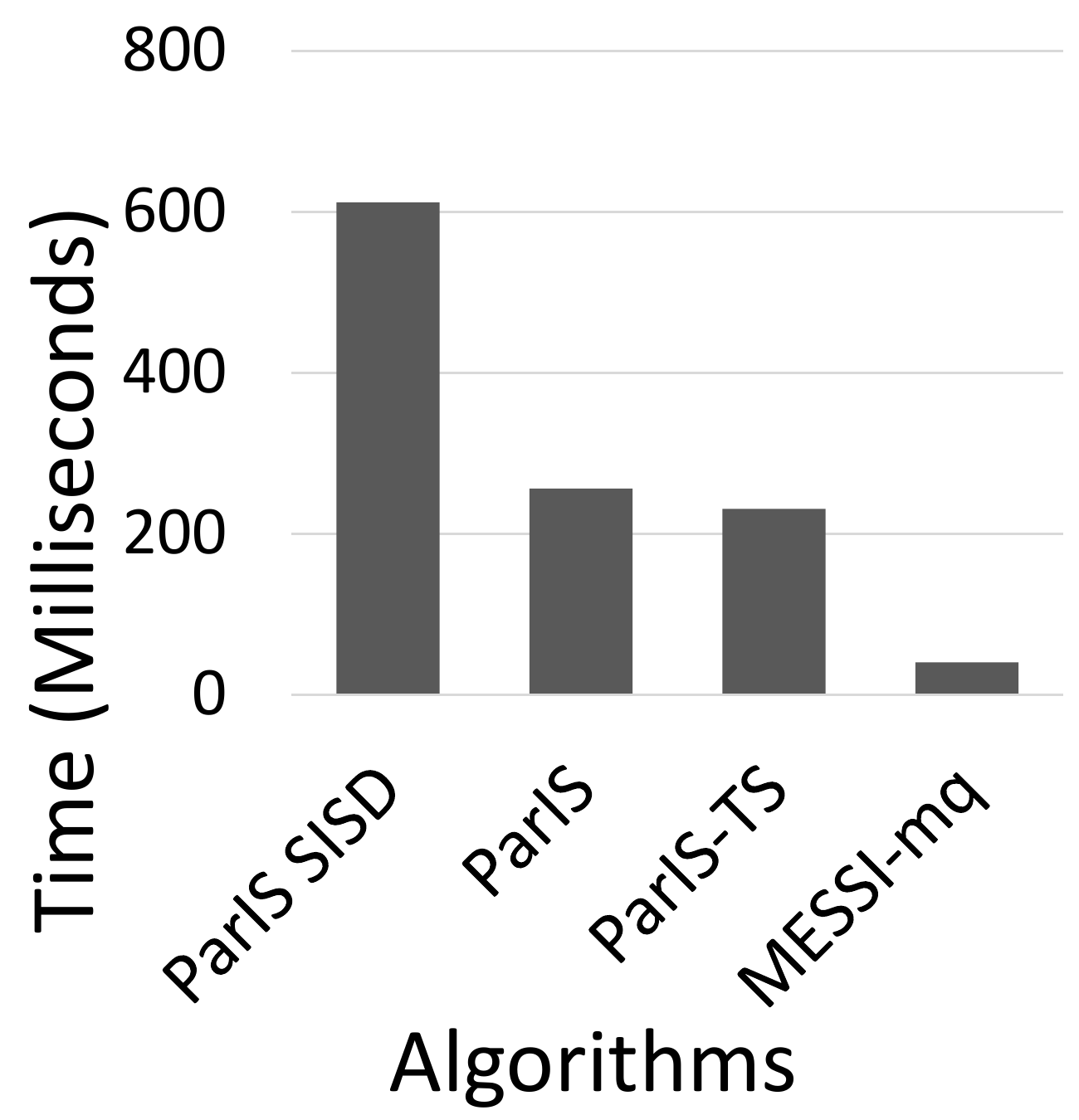}
		\caption{{Query answering performance benefit breakdown}} 
		\label{fig:stepbystepquery}
	\end{minipage}
\hspace*{0.3cm}
		\begin{minipage}[b]{0.5\columnwidth}
		\includegraphics[width=\columnwidth]{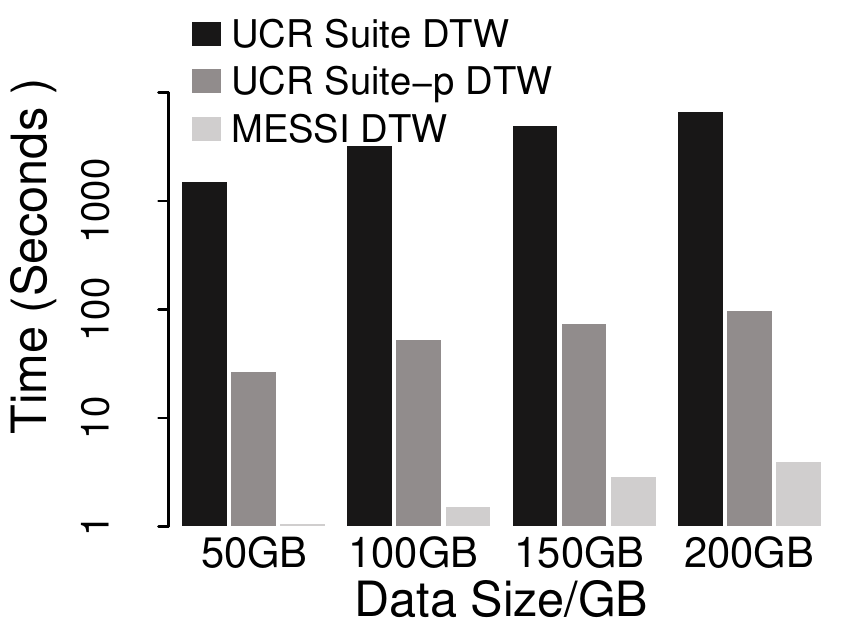}
	\caption{MESSI query answering time for DTW distance (synthetic data, 10\% warping window)}
	\label{fig:dtw}
\end{minipage}
\end{figure}


%% file: relatedwork.tex
\section{Related Work}
\label{sec:rel}

Various dimensionality reduction techniques exist 
for data series, which can then 
be scanned and filtered~\cite{DBLP:conf/kdd/KashyapK11,Li1996} or indexed and pruned~\cite{Guttman1984, Assent2008, wang2013data, shieh2008sax,Shieh2009, zoumpatianos2016ads, peng2018paris,  DBLP:journals/pvldb/KondylakisDZP18, ulissevldb}
during query answering.
We follow the same approach of indexing the series based on their summaries,
though our work is the first to exploit the parallelization opportunities offered by modern hardware, 
in order to accelerate in-memory index construction and similarity search for data series. 
The work closest to ours is ParIS~\cite{peng2018paris}, which also exploits modern hardware, but was designed for disk-resident datasets.
We discussed this work in more detail in Section~\ref{sec:prelim}.

FastQuery is an approach used to accelerate search operations in scientific data~\cite{chou2011fastquery}, 
based on the construction of bitmap indices.
In essence, the iSAX summarization used in our approach is an equivalent solution, 
though, specifically designed for sequences (which have high dimensionalities).

The interest in using SIMD instructions for improving the performance of data management solutions is not new~\cite{zhou2002implementing}.
However, it is only more recently that relatively complex algorithms were extended in order to take advantage of this hardware characteristic. 
Polychroniou et al.~\cite{polychroniou2015rethinking}  introduced design principles for efficient vectorization of in-memory database operators (such as selection scans, hash tables, and partitioning). 
For data series in particular, previous work has used SIMD for Euclidean distance computations~\cite{tang2016exploit}. 
Following~\cite{peng2018paris}, in our work we use SIMD both for the computation of Euclidean distances, 
as well as for the computation of lower bounds, 
which involve branching operations.

Multi-core CPUs offer thread parallelism through multiple cores and simultaneous multi-threading (SMT).
Thread-Level Parallelism (TLP) methods, 
like multiple independent cores and hyper-threads are used to increase efficiency~\cite{gepner2006multi}.

A recent study proposed a high performance temporal index similar to time-split B-tree (TSB-tree), 
called TSBw-tree, which focuses on transaction time databases~\cite{LometN15}. 
Binna et al.~\cite{binna2018hot}, present the Height Optimized Trie (HOT), 
a general-purpose index structure for main-memory database systems, while Leis et al.~\cite{leis2013adaptive} describe an in-memory adaptive Radix indexing technique that is designed for modern hardware. 
Xie et al.~\cite{xie2018comprehensive}, 
study and analyze five recently proposed indices, 
i.e., FAST, Masstree, BwTree, ART and PSL 
and identify the effectiveness of common optimization techniques, 
including hardware dependent features such as SIMD, NUMA and HTM. 
They argue that there is no single optimization strategy that fits all situations, due to the differences in the dataset and workload characteristics. 
Moreover, they point out the significant performance gains that the exploitation of modern hardware features, 
such as SIMD processing and multiple cores bring to in-memory indices. 

We note that the indices described above are not suitable for data series (that can be thought of as high-dimensional data), which is the focus of our work, and which pose very specific data management challenges with their hundreds, or thousands of dimensions (i.e., the length of the sequence).

%


Techniques specifically designed for modern hardware and in-memory operation have also been studied in the context of adaptive indexing~\cite{alvarez2014main}, and data mining~\cite{tatikonda2008adaptive}.

%

%% file: conclusions.tex
\section{Conclusions} 
\label{sec:conclusions}

\ignore{
Data series are a very common data type, with increasingly larger collections being generated by applications in many and diverse domains.
In many exploration and analysis pipelines, similarity search is a key operation, which is nevertheless challenging to efficiently support over large data series collections.

In this work, }
We proposed MESSI, 
a data series index designed for in-memory operation by 
exploiting the parallelism opportunities of modern hardware. 
MESSI is up to 4x faster in index construction and up to 11x faster in query answering than the state-of-the-art solution, and is the first technique to answer exact similarity search queries on 100GB datasets in $\sim$50msec.
This 
level of 
performance enables for the first time interactive data exploration on very large data series collections.



\vspace*{0.20cm}\noindent{\bf Acknowledgments}
Work supported by Chinese Scholarship Council, FMJH Program PGMO, EDF, Thales and HIPEAC 4.
Part of work performed while P. Fatourou was visiting LIPADE, and while B. Peng was visiting CARV, FORTH ICS.

%% file: mainvs.bbl
\begin{thebibliography}{10}
\providecommand{\url}[1]{#1}
\csname url@samestyle\endcsname
\providecommand{\newblock}{\relax}
\providecommand{\bibinfo}[2]{#2}
\providecommand{\BIBentrySTDinterwordspacing}{\spaceskip=0pt\relax}
\providecommand{\BIBentryALTinterwordstretchfactor}{4}
\providecommand{\BIBentryALTinterwordspacing}{\spaceskip=\fontdimen2\font plus
\BIBentryALTinterwordstretchfactor\fontdimen3\font minus
  \fontdimen4\font\relax}
\providecommand{\BIBforeignlanguage}[2]{{%
\expandafter\ifx\csname l@#1\endcsname\relax
\typeout{** WARNING: IEEEtran.bst: No hyphenation pattern has been}%
\typeout{** loaded for the language `#1'. Using the pattern for}%
\typeout{** the default language instead.}%
\else
\language=\csname l@#1\endcsname
\fi
#2}}
\providecommand{\BIBdecl}{\relax}
\BIBdecl

\bibitem{DBLP:journals/sigmod/Palpanas15}
T.~Palpanas, ``Data series management: The road to big sequence analytics,''
  \emph{{SIGMOD} Record}, 2015.

\bibitem{fulfillingtheneed}
K.~Zoumpatianos and T.~Palpanas, ``Data series management: Fulfilling the need
  for big sequence analytics,'' in \emph{ICDE}, 2018.

\bibitem{Palpanas2019}
T.~Palpanas and V.~Beckmann, ``Report on the first and second interdisciplinary
  time series analysis workshop (itisa),'' \emph{SIGMOD Rec.}, "Accepted for
  publication, 2019.

\bibitem{lernaeanhydra}
K.~Echihabi, K.~Zoumpatianos, T.~Palpanas, and H.~Benbrahim, ``The lernaean
  hydra of data series similarity search: An experimental evaluation of the
  state of the art,'' \emph{{PVLDB}}, 2018.

\bibitem{lernaeanhydra2}
------, ``Return of the lernaean hydra: Experimental evaluation of data series
  approximate similarity search,'' \emph{{PVLDB}}, 2019.

\bibitem{zoumpatianos2016ads}
K.~Zoumpatianos, S.~Idreos, and T.~Palpanas, ``Ads: the adaptive data series
  index,'' \emph{{VLDB} J.}, 2016.

\bibitem{peng2018paris}
B.~Peng, T.~Palpanas, and P.~Fatourou, ``Paris: The next destination for fast
  data series indexing and query answering,'' \emph{IEEE BigData}, 2018.

\bibitem{Fekete:2016}
J.-D. Fekete and R.~Primet, ``Progressive analytics: A computation paradigm for
  exploratory data analysis,'' \emph{CoRR}, 2016.

\bibitem{Airbus}
A.~Guillaume, ``{Head of Operational Intelligence Department Airbus. Personal
  communication.}'' 2017.

\bibitem{rakthanmanon2011}
T.~Rakthanmanon, E.~J. Keogh, S.~Lonardi, and S.~Evans, ``Time series
  epenthesis: Clustering time series streams requires ignoring some data,'' in
  \emph{ICDM}, 2011, pp. 547--556.

\bibitem{Shieh2009}
J.~Shieh and E.~Keogh, ``{iSAX: disk-aware mining and indexing of massive time
  series datasets},'' \emph{DMKD}, no.~1, 2009.

\bibitem{Shandola2009}
V.~Chandola, A.~Banerjee, and V.~Kumar, ``Anomaly detection: A survey,''
  \emph{CSUR}, 2009.

\bibitem{DBLP:journals/datamine/MueenKZCWS11}
A.~Mueen, E.~J. Keogh, Q.~Zhu, S.~Cash, M.~B. Westover, and N.~B. Shamlo, ``A
  disk-aware algorithm for time series motif discovery,'' \emph{DAMI}, 2011.

\bibitem{Agrawal1993}
R.~Agrawal, C.~Faloutsos, and A.~N. Swami, ``Efficient similarity search in
  sequence databases,'' in \emph{FODO}, 1993.

\bibitem{rakthanmanon2012searching}
T.~Rakthanmanon, B.~J.~L. Campana, A.~Mueen, G.~E. A. P.~A. Batista, M.~B.
  Westover, Q.~Zhu, J.~Zakaria, and E.~J. Keogh, ``Searching and mining
  trillions of time series subsequences under dynamic time warping,'' in
  \emph{SIGKDD}, 2012.

\bibitem{shieh2008sax}
J.~Shieh and E.~Keogh, ``i sax: indexing and mining terabyte sized time
  series,'' in \emph{SIGKDD}, 2008.

\bibitem{wang2013data}
Y.~Wang, P.~Wang, J.~Pei, W.~Wang, and S.~Huang, ``A data-adaptive and dynamic
  segmentation index for whole matching on time series,'' \emph{VLDB}, 2013.

\bibitem{isax2plus}
A.~Camerra, J.~Shieh, T.~Palpanas, T.~Rakthanmanon, and E.~Keogh, ``{Beyond One
  Billion Time Series: Indexing and Mining Very Large Time Series Collections
  with iSAX2+},'' \emph{KAIS}, vol.~39, no.~1, 2014.

\bibitem{MueenNL10}
A.~Mueen, S.~Nath, and J.~Liu, ``Fast approximate correlation for massive
  time-series data,'' in \emph{SIGMOD}, 2010.

\bibitem{lomont2011introduction}
C.~Lomont, ``Introduction to intel advanced vector extensions,'' \emph{Intel
  White Paper}, 2011.

\bibitem{tang2016exploit}
B.~Tang, M.~L. Yiu, Y.~Li \emph{et~al.}, ``Exploit every cycle: Vectorized time
  series algorithms on modern commodity cpus,'' in \emph{IMDM}, 2016.

\bibitem{keogh2001dimensionality}
E.~Keogh, K.~Chakrabarti, M.~Pazzani, and S.~Mehrotra, ``Dimensionality
  reduction for fast similarity search in large time series databases,''
  \emph{KAIS}, 2001.

\bibitem{DBLP:journals/pvldb/KondylakisDZP18}
H.~Kondylakis, N.~Dayan, K.~Zoumpatianos, and T.~Palpanas, ``Coconut: {A}
  scalable bottom-up approach for building data series indexes,''
  \emph{{PVLDB}}, 2018.

\bibitem{ulissevldb}
M.~Linardi and T.~Palpanas, ``Scalable, variable-length similarity search in
  data series: The ulisse approach,'' \emph{{PVLDB}}, 2019.

\bibitem{gogolou2019progressive}
A.~Gogolou, T.~Tsandilas, T.~Palpanas, and A.~Bezerianos, ``Progressive
  similarity search on time series data,'' in \emph{{EDBT}}, 2019.

\bibitem{sourcescode}
http://helios.mi.parisdescartes.fr/~themisp/messi/, 2019.

\bibitem{yi2000fast}
B.-K. Yi and C.~Faloutsos, ``Fast time sequence indexing for arbitrary lp
  norms,'' in \emph{VLDB}.\hskip 1em plus 0.5em minus 0.4em\relax Citeseer,
  2000.

\bibitem{iris}
``{Incorporated Research Institutions for Seismology} -- {Seismic Data
  Access},'' http://ds.iris.edu/data/access/, 2016.

\bibitem{url:SALD}
``Southwest university adult lifespan dataset (sald),''
  \url{http://fcon_1000.projects.nitrc.org/indi/retro/sald.html}, 2018.

\bibitem{berndt1994using}
D.~J. Berndt and J.~Clifford, ``Using dynamic time warping to find patterns in
  time series.'' in \emph{AAAIWS}, 1994.

\bibitem{keogh2005exact}
E.~Keogh and C.~A. Ratanamahatana, ``Exact indexing of dynamic time warping,''
  \emph{Knowledge and information systems}, 2005.

\bibitem{DBLP:conf/kdd/KashyapK11}
S.~Kashyap and P.~Karras, ``Scalable knn search on vertically stored time
  series,'' in \emph{SIGKDD}, 2011, pp. 1334--1342.

\bibitem{Li1996}
C.~Li, P.~S. Yu, and V.~Castelli, ``Hierarchyscan: {A} hierarchical similarity
  search algorithm for databases of long sequences,'' in \emph{ICDE}, 1996.

\bibitem{Guttman1984}
A.~Guttman, ``R-trees: {A} dynamic index structure for spatial searching,'' in
  \emph{SIGMOD}, 1984, pp. 47--57.

\bibitem{Assent2008}
I.~Assent, R.~Krieger, F.~Afschari, and T.~Seidl, ``The ts-tree: efficient time
  series search and retrieval,'' in \emph{{EDBT}}, 2008.

\bibitem{chou2011fastquery}
J.~Chou, K.~Wu \emph{et~al.}, ``Fastquery: A parallel indexing system for
  scientific data,'' in \emph{CLUSTER}.\hskip 1em plus 0.5em minus 0.4em\relax
  IEEE, 2011, pp. 455--464.

\bibitem{zhou2002implementing}
J.~Zhou and K.~A. Ross, ``Implementing database operations using simd
  instructions,'' in \emph{SIGMOD}.\hskip 1em plus 0.5em minus 0.4em\relax ACM,
  2002.

\bibitem{polychroniou2015rethinking}
O.~Polychroniou, A.~Raghavan, and K.~A. Ross, ``Rethinking simd vectorization
  for in-memory databases,'' in \emph{SIGMOD}.\hskip 1em plus 0.5em minus
  0.4em\relax ACM, 2015.

\bibitem{gepner2006multi}
P.~Gepner and M.~F. Kowalik, ``Multi-core processors: New way to achieve high
  system performance,'' in \emph{PAR ELEC}, 2006.

\bibitem{LometN15}
D.~B. Lomet and F.~Nawab, ``High performance temporal indexing on modern
  hardware,'' in \emph{{ICDE}}, 2015.

\bibitem{binna2018hot}
R.~Binna, E.~Zangerle, M.~Pichl, G.~Specht, and V.~Leis, ``Hot: A height
  optimized trie index for main-memory database systems,'' in
  \emph{SIGMOD}.\hskip 1em plus 0.5em minus 0.4em\relax ACM, 2018.

\bibitem{leis2013adaptive}
V.~Leis, A.~Kemper, and T.~Neumann, ``The adaptive radix tree: Artful indexing
  for main-memory databases.'' in \emph{ICDE}, 2013.

\bibitem{xie2018comprehensive}
Z.~Xie, Q.~Cai, G.~Chen, R.~Mao, and M.~Zhang, ``A comprehensive performance
  evaluation of modern in-memory indices,'' in \emph{ICDE}, 2018.

\bibitem{alvarez2014main}
V.~Alvarez, F.~M. Schuhknecht, J.~Dittrich, and S.~Richter, ``Main memory
  adaptive indexing for multi-core systems,'' in \emph{DaMoN}, 2014.

\bibitem{tatikonda2008adaptive}
S.~Tatikonda and S.~Parthasarathy, ``An adaptive memory conscious approach for
  mining frequent trees: implications for multi-core architectures,'' in
  \emph{SIGPLAN}.\hskip 1em plus 0.5em minus 0.4em\relax ACM, 2008.

\end{thebibliography}
